\documentclass[final]{siamltex}
\usepackage{graphicx,amsfonts,amsmath,cite,url}
\usepackage{color}
\usepackage{psfrag}

\begin{document}

\title{Comparing Community Structure to Characteristics\\ in Online Collegiate Social Networks}

\author{Amanda L. Traud$^{1,2}$, Eric D. Kelsic$^{3}$, Peter J. Mucha$^{1,4}$,\\ and Mason A. Porter$^{5,6}$\\
  $^1$\footnotesize{Carolina Center for Interdisciplinary Applied Mathematics, Department of Mathematics,\\ University of North Carolina, Chapel Hill, NC 27599-3250, USA} \\
  $^2$\footnotesize{Carolina Population Center,\\ University of North Carolina, Chapel Hill, NC 27516-2524, USA} \\
  $^3$\footnotesize{Department of Systems Biology, Harvard Medical School, \\ Harvard University, Boston, MA 02115, USA} \\
  $^4$\footnotesize{Institute for Advanced Materials, Nanoscience \& Technology,\\
  University of North Carolina, Chapel Hill, NC 27599-3216, USA} \\
  $^5$\footnotesize{Oxford Centre for Industrial and Applied Mathematics, Mathematical Institute, University of Oxford, OX1 3LB, UK}\\
  $^6$\footnotesize{CANDyN Complexity Centre, University of Oxford, OX1 1HB, UK}\\
}
\maketitle


\section*{Abstract}

\emph{We study the structure of social networks of students by examining the graphs of Facebook ``friendships''  at five American universities at a single point in time.  We investigate each single-institution network's community structure and employ graphical and quantitative tools, including standardized pair-counting methods, to measure the correlations between the network communities and a set of self-identified user characteristics (residence, class year, major, and high school).  We review the basic properties and statistics of the pair-counting indices employed and recall, in simplified notation, a useful analytical formula for the $z$-score of the Rand coefficient.  Our study illustrates how to examine different instances of social networks constructed in similar environments, emphasizes the array of social forces that combine to form ``communities,'' and
leads to comparative observations about online social lives that can be used to infer comparisons about offline social structures.  In our illustration of this methodology, we calculate the relative contributions of different characteristics to the community structure of individual universities and subsequently compare these relative contributions at different universities, measuring for example the importance of common high school affiliation to large state universities and the varying degrees of influence common major can have on the social structure at different universities. The heterogeneity of communities that we observe indicates that these networks typically have multiple organizing factors rather than a single dominant one.
}


\section{Introduction} \label{intro}


Social networks are a ubiquitous part of everyday life. Although they have long been studied by social scientists \cite{faust}, the mainstream awareness of their ubiquity has arisen only recently, in part because of the rise of social networking sites (SNSs) on the World Wide Web. Since their introduction, SNSs such as Friendster, MySpace, Facebook, Orkut, LinkedIn, and hundreds of others have attracted hundreds of millions of users, many of whom have integrated SNSs into their daily lives to communicate with friends, send e-mails, solicit opinions or votes, organize events, spread ideas, find jobs, and more \cite{boydell}.  Facebook, an SNS launched in February 2004, now overwhelms numerous aspects of everyday life, having become an especially popular obsession among college and high school students (and, increasingly, among other members of society) \cite{boydheart,boydell,lewis,oldboy}.  Facebook members can create self-descriptive profiles that include links to the profiles of their ``friends," who may or may not be offline friends.  Facebook requires that anybody who one wants to add as a friend confirm the relationship, so Facebook friendships define a network (graph) of reciprocated ties (undirected edges) that connect individual users.

The global organization of real-world networks typically includes coexisting modular (horizontal) and hierarchical (vertical) organizational structures \cite{ourreview,santobig,Newman2003,newman2010,cald}.  Myriad papers have attempted to interpret such organization through the computation of structural modules or \textit{communities} \cite{ourreview,santobig}, which are defined in terms of mesoscopic groups of nodes with more internal connections (between nodes in the group) than external connections (between nodes in the group and nodes in other groups).   Such communities, which are not typically identified in advance, are often considered to not be merely structural modules but are also expected to have functional importance because of the large number of common ties among nodes in a community.  Additionally, prior empirical studies have observed some correspondence between communities and ``ground truth" groups in social and biological networks
\cite{ourreview}.  
For example, communities in social networks might correspond to circles of friends or business associates, communities in the World Wide Web might encompass pages on closely-related topics, communities in metabolic networks have been used to find functional modules \cite{amaral}, and communities have been used to identify and measure political polarization in legislative processes in the U.S.~Congress \cite{waugh,yan}.

As discussed at length in two recent review articles \cite{ourreview,santobig} and references therein, the classes of techniques available to detect communities are both numerous and diverse; they include hierarchical clustering methods such as single linkage clustering, centrality-based methods, local methods, optimization of quality functions such as modularity and similar quantities, spectral partitioning, likelihood-based methods, and more.  In addition to remarkable successes on benchmark examples, investigations of community structure have led to success stories in diverse application areas---including the reconstruction of college football conferences \cite{structpnas} and the investigation of such structures in algorithmic rankings \cite{bcs}; the analysis of committee assignments \cite{congshort}, legislation cosponsorship \cite{yan}, and voting blocs \cite{waugh} in the U.S.~Congress; the examination of functional groups in metabolic networks \cite{amaral}; the study of ethnic preferences in school friendship networks \cite{marta}; and the study of social structures in mobile-phone conversation networks \cite{jp}.

In this paper, we investigate the community structures of complete Facebook networks whose links represent reciprocated ``friendships'' between user pages (nodes) within each of five American universities during a single-time snapshot in September 2005.  Our primary aim in this paper is to use an unsupervised algorithm to compute the community structure---consisting of clusters of nodes---of these universities and to determine how well the demographic labels included in the data correspond to algorithmically computed clusters.  We consider only ties between students at the same institution, yielding five separate realizations of university social networks and allowing us to compare the structures at different institutions.

The rest of this paper is organized as follows. In Section \ref{sec:methods}, we describe our principal methods: the employed community-detection method, visual exploration of identified communities, and standardized pair-counting methods for quantitative comparison of communities with demographic data.  We present more details about the data in Section \ref{sec:data}. We then describe and discuss the results that we obtained for the five institutions in Section \ref{sec:examples} before concluding in Section \ref{sec:discussion}.


\section{Comparing Communities} \label{sec:methods}

\begin{figure}[htbp]
\centerline{
\includegraphics[width=0.45\textwidth]{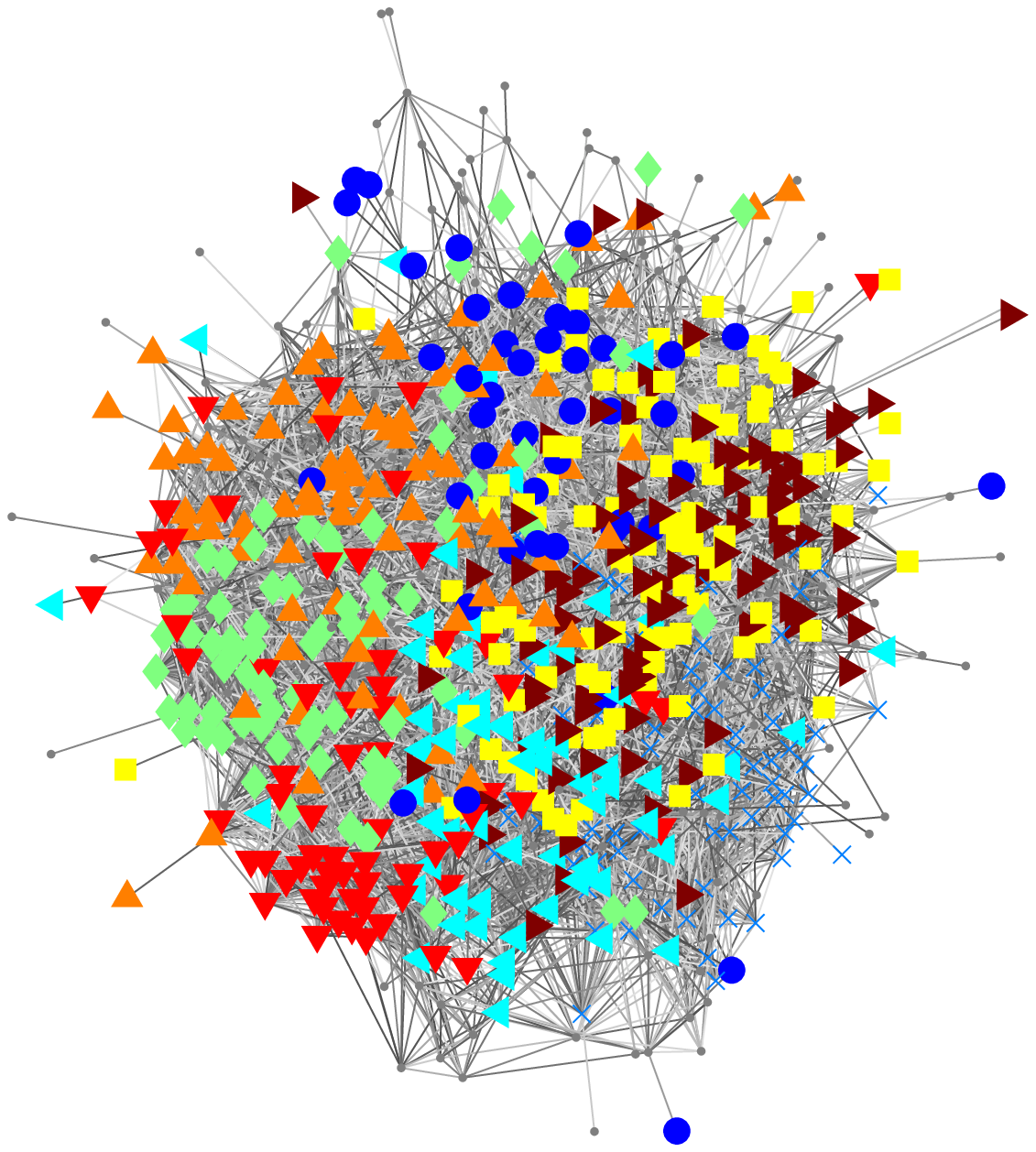}
\includegraphics[width=0.55\textwidth]{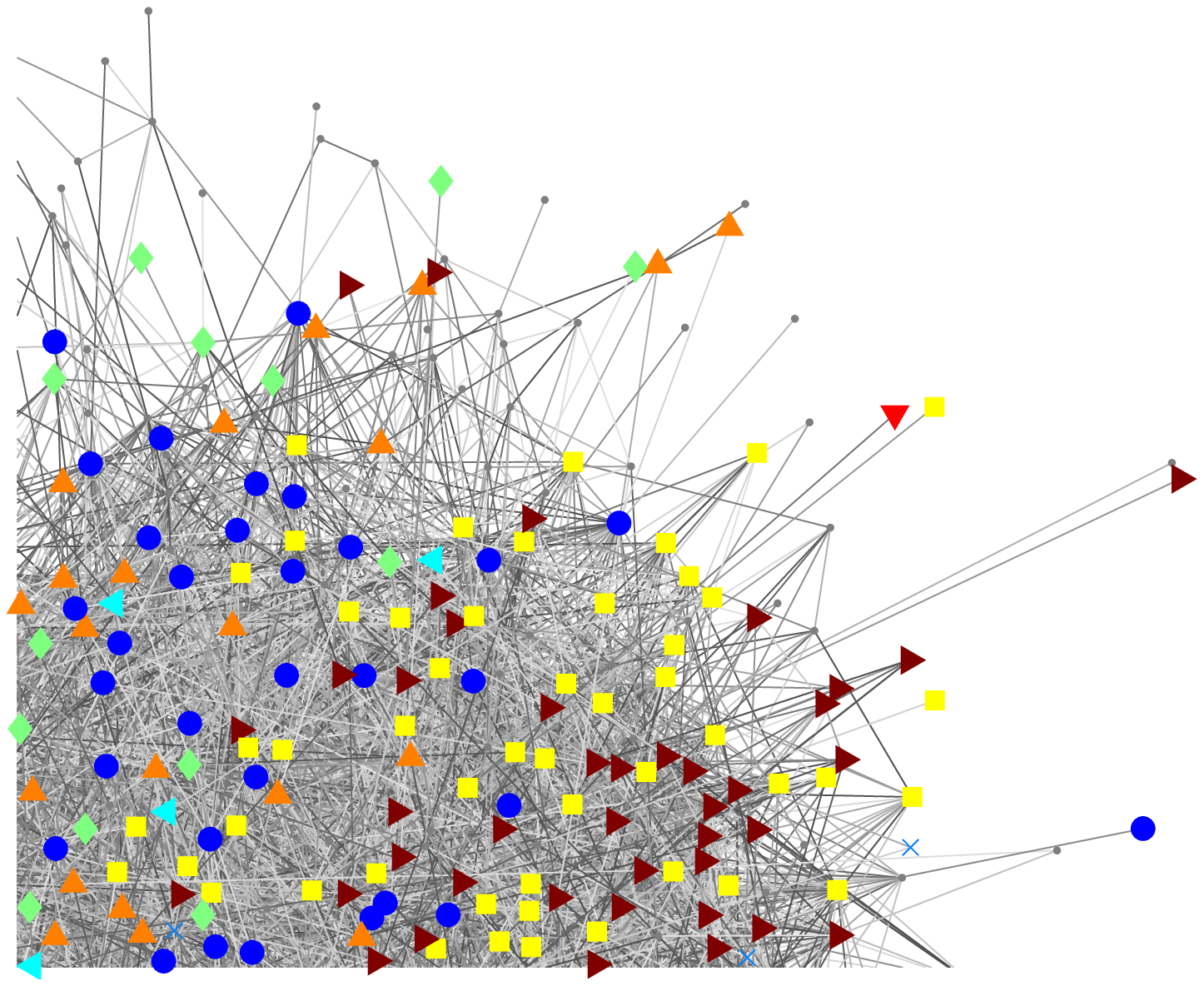}
		}
\caption{[Color] (Left) A Fruchterman-Reingold visualization \cite{FruchtermanReingold91} of the largest connected component of the Caltech Facebook network. Node shapes and colors indicate House affiliation (gray dots denote users who did not identify an affiliation), and the edges are randomly shaded for easy viewing. (Right) Magnification of a portion of the network.  Clusters of nodes with the same color/shape suggest that House affiliation affects the existence of friendships/edges.}
\label{kkexample}
\end{figure}

A social network with a single type of connection between nodes can be represented as an adjacency matrix $A$ whose elements $A_{ij}$ give the weight of the tie between nodes $i$ and $j$. The Facebook networks we study are unweighted, so $A_{ij} \in \{0,1\}$, where the value is $1$ if a tie exists and $0$ if it does not.  The resulting tangle of nodes and links, which we show for the California Institute of Technology (Caltech) Facebook network in Fig.~\ref{kkexample}, can obfuscate any organizational structure that might be present.

One approach to analyzing such data is to employ exponential random graph models (see, e.g., \cite{RobinsERGM}), statistically fitting an underlying model for the presence of links.  While such models (which can incorporate local network features) are potentially valuable for understanding the microscopic processes that underly the links between individual nodes, we take a different approach, focusing on groups of friends that form structural ``communities''---groups of nodes that contain more internal connections (links between nodes in the group) than external connections (between nodes of the group and nodes in other groups) \cite{ourreview,santobig}.  Our approach is motivated in part by the features of the Caltech data (discussed in detail in Sections \ref{sec:data} and \ref{sec:examples}).  Although precise results obviously vary from one model specification to another, performing a logistic regression on the dyads (pairs of nodes) yields comparable coefficients for link presence between users from the same House as from the same high school. However, there are significantly more users sharing the former than the latter at Caltech. While common high school is unsurprisingly important at the dyadic level (in the rare cases that it happens), common House affiliation is apparently much more important for understanding structures that consist of larger groups of individuals. Accordingly, our goal in this section is to discuss how to compare the composition of algorithmically-determined  communities to groups defined based on common user characteristics.

We identify communities using spectral optimization \cite{newman2006pre} (followed by supplementary Kernighan-Lin node-swapping steps \cite{kl}) of the ``modularity'' quality function $Q = \sum_i (e_{ii} - b_i^2)$, where $e_{ij}$ denotes the fraction of ends of edges in group~$i$ for which the other end of the edge lies in group~$j$ and $b_i = \sum_j e_{ij}$ is the fraction of all ends of edges that lie in group~$i$. High values of modularity correspond to community assignments with greater numbers of intra-community links than expected at random (with respect to a particular null model \cite{newman2006pre,ourreview,santobig}).  Numerous other community detection methods are also available.  However, our focus in the present paper is on studying communities after they are obtained, and our methods can be applied to the output of any community-detection algorithm in which each node is assigned to precisely one community.  Such an assignment of nodes to communities constitutes a partition of the original graph.  We seek a means to compare an algorithmically-obtained partition to partitions based on information that we have about Facebook user characteristics---class year, dormitory (House), high school,
and major---as a means of exploring the roles of such characteristics in the social structures of each institution.  An online social network is an imperfect proxy for an offline network, but our comparisons are nevertheless expected to yield interesting insights about the social life at the universities we study.



\subsection{Visual Comparisons} \label{subsec:visual}

\begin{figure}[ht]
    \centerline{
        \includegraphics[width=0.45\textwidth]{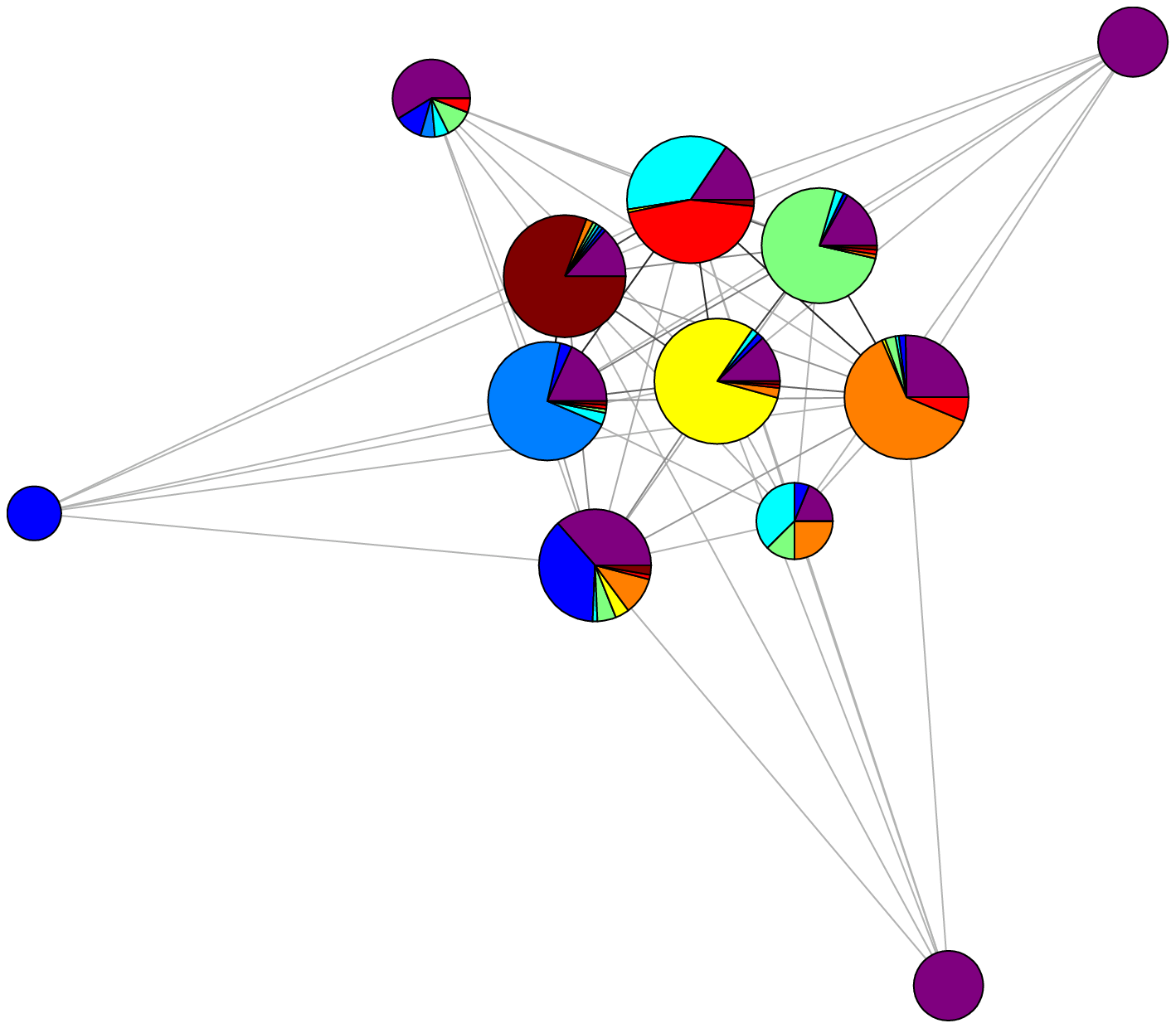}
        \includegraphics[width=0.5\textwidth]{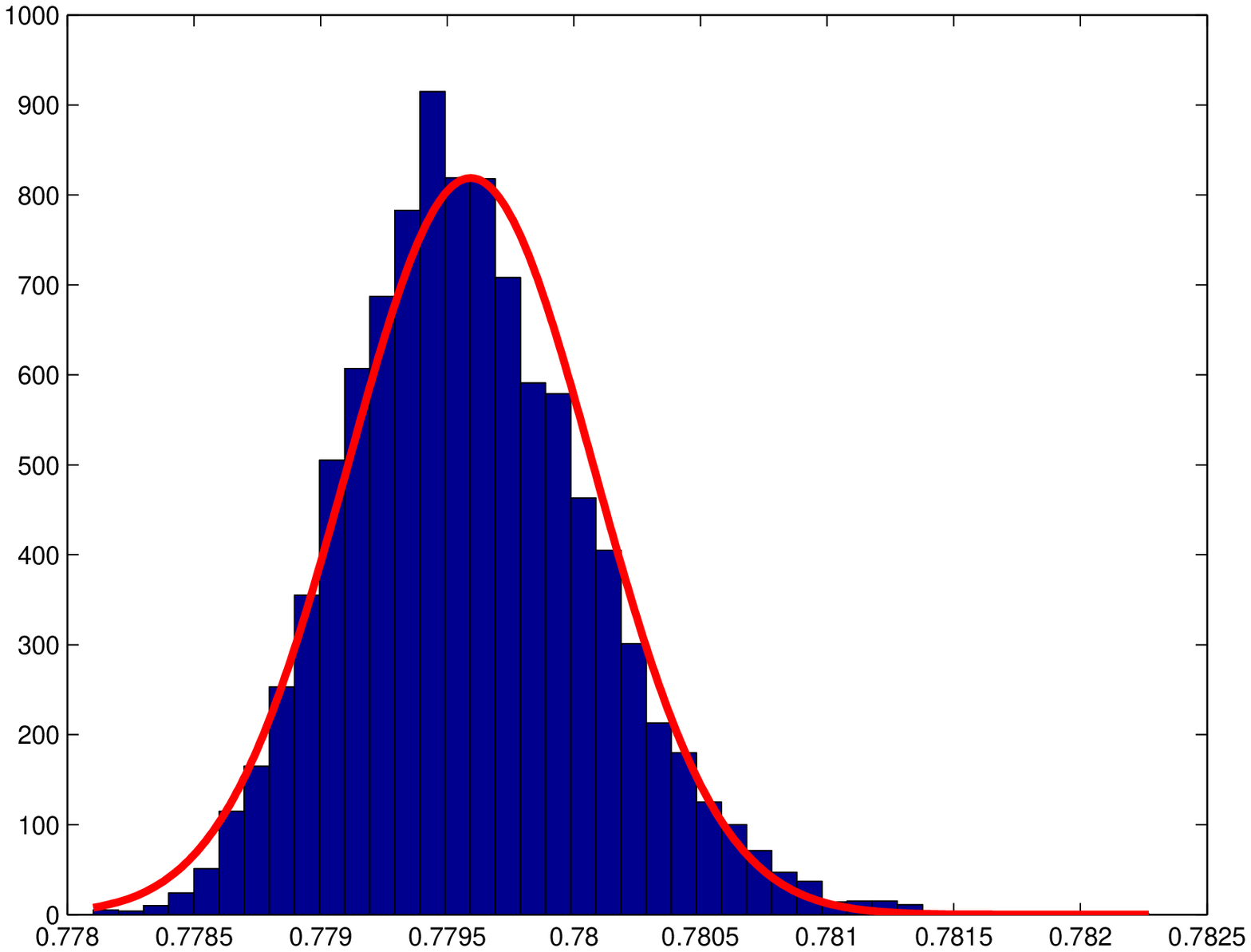}
}
    \caption{[Color] (Left) Force-directed layout of Caltech communities, each represented by a pie chart with area proportional to population and colored by House affiliation (with purple signifying missing information).  (Right) Distribution of Rand coefficients comparing these $12$ Caltech communities with random permutations of partitions into 9 House categories (including ``Missing'').  For comparison, we plot in red a Gaussian with the sample mean and variance. As our smallest data set, this yields the most extreme deviation from the Gaussian in our permutation tests.
    }
    \label{fig:Caltech}
\end{figure}

The demographic composition of communities is sometimes clear from visual inspection. This is the case with the community structure of the Caltech network, which agrees closely with its undergraduate ``House'' system.  In Fig.~\ref{fig:Caltech}, we show a force-directed layout of Caltech's 12 communities (yielding a modularity of $Q \doteq 0.4002$), which we show as pies with area proportional to the number of constituent nodes. Purple slices signify individuals who did not identify a House affiliation.

Unlike other universities (see Section \ref{sec:examples}), we find that House affiliation is the primary organizing principle of the communities in the Caltech network, which is what we expected because Caltech's House structure is so dominant socially. Indeed, each pie in Fig.~\ref{fig:Caltech} is dominated by members of one House. Moreover, many pies include a significant number of people who identify ``Avery House" as their affiliation (dark blue), which is expected because of its different residency rules (members of all Houses could live in Avery at the time of this data). Given the promotion of Avery House to official House status after our data snapshot, it is natural to wonder if community detection on current data would find a community dominated by Avery.  Investigating the formation of such a community using longitudinal data would be even more interesting, but is beyond the scope of our data. In principle, one can also make limited predictions based on the compositions of the communities about users who did not volunteer their House affiliation.

Despite this demonstration of the utility of visualizing communities, it is typically necessary to perform quantitative analyses after detecting communities, as Caltech is unusual among universities in having a single characteristic that aligns so closely with its communities.  For other institutions, we observe more heterogeneous communities, and it is typically difficult to visually assess which characteristics best correlate with the communities or even whether there is any strong correlation at all. To investigate the social organization of communities at such universities, it is thus essential to quantitatively compare the detected communities with the available demographic groups.  Such considerations apply broadly to community detection in most networks \cite{ourreview}.


\subsection{Pair Counting}

As discussed in Refs.~\cite{meila,newman2008pre}, methods to compare graph partitions can be classified roughly into three groups: (1) pair counting, (2) cluster matching, and (3) information-theoretic techniques. Cluster matching might be particularly problematic in the present context, as the numbers and sizes of groups vary significantly across the comparisons, which makes the essential identifications across partitions rather difficult. We focus on a collection of pair-counting methods, in part because of their convenient algebraic description, as one just needs to count the ways that pairs of nodes are grouped across two partitions. That same simplicity can also be a weakness, as it can present a serious interpretation difficulty because of the unclear range of ``good'' scores.  However, as we will show in Section \ref{sec:zscores}, standardization of pair-counting scores provides a unified interpretation of a number of seemingly disparate pair-counting measures and is particularly useful for the present setting. We also compare these results with those obtained using variation of information (VI) \cite{meila}.


A pair-counting method defines a similarity score by counting each pair of nodes drawn from the $n$ nodes of a network according to whether the pair falls in the same or in different groups in each partition. Pair-counting methods comprise a subset of a more general class of association measures that can be used for studying unordered (i.e., categorical) contingency tables \cite{HubertArabie,Kulisnkaya,meila}.  We denote the counts of node pairs in each classification as $w_{11}$ (pairs classified together in both partitions), $w_{10}$ (same in the first but different in the second), $w_{01}$ (different in the first but same in the second), and $w_{00}$ (different in both).  The sum of these quantities is, by definition, equal to the total number $M$ of node pairs: $M = w_{11}+w_{10}+w_{01}+w_{00}=\binom{n}{2}=n(n-1)/2$.  Given two partitions of a network, one can obtain many different pair-counting similarity coefficients using different algebraic combinations of the $w_{\alpha\beta}$ counts.

We first consider the Rand similarity coefficient $S_\mathrm{R} = (w_{11}+w_{00})/M$ \cite{Rand71}, which counts the fraction of node pairs identified the same way by both
partitions (either together in both or separate in both). Bounded between $0$ (no
similar pair placements) and $1$ (identical partitions), the Rand coefficient is extremely intuitive and can be used fruitfully in many settings. However, it has an important deficiency: The Rand coefficient for two network partitions that each contain large numbers of categories is skewed towards the value $1$ because of the large fraction of node pairs that are placed in different groups even when comparing two partitions with little in common.

If one wishes to exclude $w_{00}$ from having an explicit role, one can use the Jaccard
index $S_\mathrm{J}=w_{11}/(w_{11}+w_{10}+w_{01})$ or the
Fowlkes-Mallows similarity coefficient $S_\mathrm{FM}=w_{11}/\sqrt{(w_{11}+w_{10})(w_{11}+w_{01})}$.  Both $S_\mathrm{J}$ and $S_\mathrm{FM}$ clearly avoid the problematic
effects of large $w_{00}$, but their ignorance of node pairs
classified similarly into different communities yields overly high values when
comparing network partitions with very few categories (or when one
partition consists of a single group).  Another index is the Minkowski coefficient $S_\mathrm{M}=\sqrt{(w_{10}+w_{01})/(w_{10}+w_{11})}$, which is asymmetric in its consideration of the two partitions.  The first serves as a distinguished reference, measuring the number of
mismatches relative to the number of similarly-grouped pairs in that
reference. Hence, $S_\mathrm{M}$ values closer to $0$ are
considered better. The $\Gamma$ similarity coefficient, defined as
\[
S_\Gamma=\frac{{Mw_{11}
 - (w_{11}  + w_{10} )(w_{11}  + w_{01} )}}{{\sqrt {(w_{11}  + w_{10} )(w_{11}  + w_{01} )(M -
(w_{11}  + w_{10} ))(M - (w_{11}  + w_{01} ))} }}\,,
\]
has the most complicated algebraic form of the similarity coefficients
that we employ.  Additional measures and discussions are available in Refs.\
\cite{Campello2007,JainDubes88,meila}.  Notably, each $S_i$ measure suffers from the difficulty of it being unclear what constitute  ``good" values, as they all depend intimately on the numbers and sizes of the groups in the partition.  (We illustrate this in Section \ref{sec:examples} with computations for the Caltech network and discuss further properties of the similarity indices in Subsection \ref{sec:zscores}.)

One can also try to alleviate the problem of identifying good similarity values by introducing various ``adjusted'' indices that report comparisons as a similarity relative to that which might be obtained at random.  For instance, one can construct adjusted indices by subtracting the expected value (under some null model, typically conditional on maintaining the numbers and sizes of groups in the two partitions) and then rescaling the result by the difference between the maximum allowed value and the mean value \cite{HubertArabie}.  One such index, using a bound on the maximum allowed value, is the Adjusted Rand coefficient \cite{HubertArabie}
\[
S_\mathrm{AR} = \frac{w_{11} - \frac{1}{M}(w_{11}+w_{10})(w_{11}+w_{01})}{\frac{1}{2}\left[(w_{11}+w_{10})+(w_{11}+w_{01})\right] - \frac{1}{M}(w_{11}+w_{10})(w_{11}+w_{01})}\,.
\]

As described in Ref.~\cite{meila}, adjusted indices can be problematic because the focus on the maximum possible values does not guarantee accurate comparisons between similarity coefficients across different settings. In particular, this implies that one cannot necessarily use similarity scores to make direct comparisons between communities and House with those between communities and high school (which is something that we specifically aim to do).  That is, even if such comparisons yield Adjusted Rand values of $0.1$ and $0.2$, it is not at all clear that the second situation should be construed to yield a closer pair of partitions than the first.  Consequently, the general problem of knowing what similarity-score values indicate a good correlation remains.



\subsection{Standardized Pair Counting} \label{sec:zscores}
\label{subsec:zscores}

Numerous studies have attempted to assess the utility of similarity measures.  However, because partitioning according to demographic traits yields a graph partitioning that typically differs significantly from that obtained using algorithmic community detection, we use a classical statistical approach, advocated in \cite{BrennanLight74,FowlkesMallows83}, wherein similarity measures are used in the context of testing significance levels of the obtained values versus those expected at random.  We recommend using a proper metric (i.e., a quantity that is a metric in the mathematical sense rather than only in an informal sense) such as variation of information \cite{meila} for comparing partitions that are close to one another.  However, in the Facebook networks, the mutual information of a pair of partitions is small compared to the total information in each. In such cases, two partitions can be relatively far from each other according to a distance measure but might nevertheless be very far in the tail of the distribution of what can be expected at random. It is consequently more appropriate to identify the pair-counting strength relative to that obtained at random, standardized by the width of the distribution via $z$-scores $z_i = (S_i-\mu_i)/\sigma_i$, which indicate the number of standard deviations $\sigma_i$ that the $S_i$-value is more correlated than the mean $\mu_i$ ($i\in\{\mathrm{FM},\Gamma,\mathrm{J},\mathrm{M},\mathrm{R},\mathrm{AR}\}$, noting the need to multiply by $-1$ for $z_\mathrm{M}$).


One can obtain $z$-scores non-parametrically using permutation tests \cite{Good2005}, though we will identify analytical formulas for $z_\mathrm{R}$ and show that the Fowlkes-Mallows, $\Gamma$, Rand, and Adjusted Rand $z$-scores are identical.  The elements $n_{ij}$ of the contingency table indicate the number of nodes that are classified into the $i$th group of the first partition and $j$th group of the second partition. As long as partitions are constrained to have the same numbers and sizes of groups as the original partitions---i.e., as long as the row and column sums, $n_{i\cdot}=\sum_j n_{ij}$ and $n_{\cdot j}=\sum_i n_{ij}$, remain constant---then the total number of pairs $M$, the number of pairs $M_1=\sum_i \binom{n_{i\cdot}}{2}$ classified the same way in the first partition, and the analogous quantity $M_2=\sum_j \binom{n_{\cdot j}}{2}$ for the second partition likewise remain constant.  This implies that any pair-counting index specified by $w_{\alpha\beta}$ counts can be equivalently specified in terms of only $w := w_{11}=\sum_{ij}\binom{n_{ij}}{2}$ because $w_{10}=M_1-w$, $w_{01}=M_2-w$, and $w_{00}=M-M_1-M_2+w$. It follows immediately that $S_\mathrm{R}$, $S_\mathrm{FM}$, $S_\Gamma$, $S_\mathrm{AR}$ are each linear functions of $w$
and hence linear functions of each other \cite{JainDubes88}.  Any similarity index $S_i$ that is a linear function of $w$ must be statistically equivalent to $w$ in any null model (given constant $M$, $M_1$, and $M_2$), with the $z$-score and $p$-value equal to that associated with the specified $w$. Meanwhile, as we demonstrate in Section \ref{sec:examples}, the $S_i$ values can have different orderings in different comparisons because of their dependence on $M$, $M_1$, and $M_2$.

It is also instructive to note the relationships between the linear-in-$w$ similarity coefficients and the Jaccard and Minkowski indices: ${1}/{S_\mathrm{J}} = -1 + ({M_1+M_2})/{w}$ and $S^2_\mathrm{M} = ({M_1+M_2-2w})/{M_1}$.  The asymmetry in the Minkowski index is clearly limited, as switching which partition is the reference changes the coefficient by a multiplicative factor. Because the square root and multiplicative inverse are both monotonic operations in the domains of these indices ($S_\mathrm{M} > 0$ and $0 \leq S_\mathrm{J} \leq 1$), it follows that the $p$-values of the cumulative distributions of each are identical to the $p$-value of $w$ itself even though the corresponding $z$-scores can be different.

In deference to the seminal presentation of the Rand index \cite{Rand71}, we refer to the $z$-score of the linear-in-$w$ scores as $z$-Rand: $z_\mathrm{R} = (w-\mu_w)/\sigma_w$, where $\mu_w$ and $\sigma_w$ are, respectively, the mean and standard deviation of $w$ (noting its equivalence by linearity to the $z$-score advocated explicitly by Brennan and Light \cite{BrennanLight74}).  In the absence of external information that indicates a need to impose specific correlations, we adopt the standard and analytically tractable assumption of a random hypergeometric distribution of equally likely assignments subject to fixed row and column sums. The expected value then becomes $\mu_w = M_1M_2/M$, as for the adjusted Rand index \cite{HubertArabie}.  The calculation of higher-order moments is more involved \cite{BrennanLight74,BrookStirling84,Hubert77,Mantel67}. In order to make $z_\mathrm{R}$ as simple as possible to calculate, we rewrite the formulas of \cite{Hubert77} as follows:
\begin{equation}\label{eq:zrand}
	z_\mathrm{R} = \frac{1}{\sigma_w}\left(w-\frac{M_1M_2}{M}\right)\,,
\end{equation}
\begin{align}
	\label{eq:varw}
	\sigma_w^2 &= \frac{M}{16} - \frac{(4M_1-2M)^2(4M_2-2M)^2}{256M^2} + \frac{C_1C_2}{16n(n-1)(n-2)}\cr &+ \frac{[(4M_1-2M)^2-4C_1-4M][(4M_2-2M)^2-4C_2-4M]}{64n(n-1)(n-2)(n-3)}\,,
\end{align}
\begin{align} \label{eq:C12}
	C_1 &= n(n^2-3n-2) - 8(n+1)M_1 + 4\sum_i n_{i\cdot}^3\,,\cr
	C_2 &= n(n^2-3n-2) - 8(n+1)M_2 + 4\sum_j n_{\cdot j}^3\,.
\end{align}

While we advocate the use of $z_\mathrm{R}$, their associated significance levels (equivalently, the $p$-values of the cumulative distribution) are not equal to those for a Gaussian distribution.  The distribution for large samples is asymptotically Gaussian \cite{Kulisnkaya}, but the distribution associated with comparing a particular pair of partitions need not be. Indeed, the tails of the distribution can be quite heavy \cite{BrookStirling84}, so the probability of obtaining extreme $z$-scores can be orders-of-magnitude higher than in the normal distribution. Nevertheless, the Gaussian approximation is frequently sufficient to gauge statistical significance (past the 95\% confidence interval). Given the straightforward calculation of \eqref{eq:zrand}--\eqref{eq:C12}, we prefer to use $z_\mathrm{R}$ directly, with the caveat that the Rand indices do not translate directly to $p$-values.

Where simple formulas for the necessary moments do not appear to be available (i.e., for the Jaccard and Minkowski indices), we resort to the computationally straightforward (albeit intensive if one desires high accuracy) method of examining distributions obtained using permutation tests \cite{Good2005}, again under the null model of equally-likely node assignments conditional on the constancy of the numbers and sizes of groups. Specifically, starting from two network partitions whose correlation we want to measure, we calculate the similarity values $S_i$ and obtain a context for these values by repeatedly computing $S_i$ under random permutation of the node assignments in one of the partitions.  (Subsequent permutation of assignments in the second partition is redundant.)  We thereby aim to compare the similarity coefficients between the two partitions to the distributions of such coefficients from the appropriate ensemble of partition pairs.  Numerical estimation of $p$-values far in the tail of the distribution (where many of our points of interest lie) necessarily requires sampling a correspondingly large number of elements. In contrast, calculating $z$-scores only requires sampling the first two moments of the distribution.  We typically use $10000$ permutations (even for the larger networks, where the number of nodes is actually larger than the number of permutations considered), confirming that the obtained $z$-scores have converged to roughly two significant figures by comparing them with those obtained using half of the permutations and also comparing $z_\mathrm{R}$ estimates with the analytical values obtained from \eqref{eq:zrand}--\eqref{eq:C12}.

\begin{figure}[ht]
    \centerline{
        \psfrag{Z Rand}[b][b]{$z_\mathrm{R}$}
        \psfrag{Z Jaccard}[bl][bl]{$z_\mathrm{J}$}
        \psfrag{Z Minkowski}[bl][bl]{$z_\mathrm{M}$}
        \psfrag{Adj. Rand}[bl][bl]{$S_\mathrm{AR}$}
        \psfrag{VI}[bl][bl]{VI}
        \psfrag{Z VI}[bl][bl]{$z_\mathrm{VI}$}
        \includegraphics[width=0.75\textwidth]{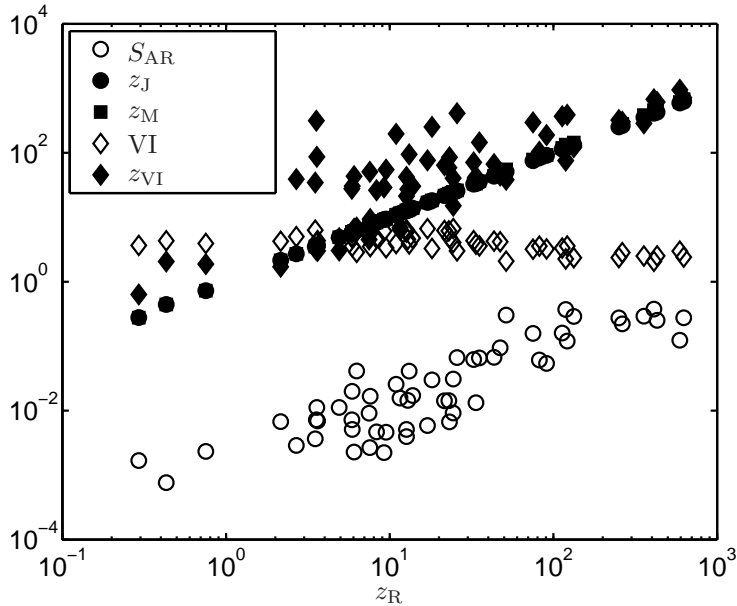}
    }
    \caption{Scatter plot of $z_{\rm R}$ (the Rand $z$-score) on the horizontal axis versus (on the vertical axis) other pair-counting $z$-scores ($z_{\rm J}$ and $z_{\rm M}$), variation of information (VI), a VI $z$-score from permutation tests, and the Adjusted Rand index $S_\mathrm{AR}$. The depicted data comes from 60 situations: algorithmically-detected communities
for the 5 universities using 4 demographic groupings and 3 networks per university (full data and gender-restricted networks of women only and men only).}
        \label{fig:ZvZR}
\end{figure}

Of course, calculating $z$-scores of the pair-counting indices is not a panacea, particularly when comparing networks of different sizes.  Nevertheless, we find them to be exceptionally useful for examining the correlations between communities and partitions by the available demographics in our Facebook data.  Before we concentrate on using these $z$-scores to measure correlations, we compare test results (similar to those discussed in Section \ref{sec:examples}) against other methods, including variation of information \cite{meila} and the (non-standardized) Adjusted Rand index $S_\mathrm{AR}$ \cite{HubertArabie} using a scatter plot versus $z_\mathrm{R}$ in Fig.~\ref{fig:ZvZR}. While $S_\mathrm{AR}$ trends positively with $z_\mathrm{R}$ (recall that $z_\mathrm{R} = z_\mathrm{AR}$), there are clearly situations
with very small $S_\mathrm{AR}$ that have much larger $z_\mathrm{R}$ values than should be expected at random.  We additionally observe that $z_\mathrm{J}$ and $z_\mathrm{M}$ each appear to be closely approximated by $z_\mathrm{R}$ at the scale of Fig.~\ref{fig:ZvZR}, though closer inspection reveals relative differences occasionally as large as 10\%.

We admit that we are questionably guilty of one of the major sins of statistical analysis, in that $z$-scores are typically a proxy for the likelihood with which one can reject an independent null hypothesis.  It is thus reasonable to question their effectiveness for the quite different task of measuring a correlation.
We stress, however, that the underlying statistic that we have standardized is a pair counting of the similarities between partitions rather than a $\chi^2$ deviation from independence.  (We note that $w$ reduces to a linear function of $\chi^2$ in the special case of uniform constant marginals \cite{BrookStirling84}.) Therefore, in the absence of enforcing a particular model for the form of the correlation between partitions, we believe this standardization of similarity scores is a reasonable way to proceed (if done so with caution).


\section{Data} \label{sec:data}

Our data, which was sent directly to us in anonymized form by Adam D'Angelo of Facebook, consists of the complete set of users (nodes) from the Facebook networks for each of five American universities and all of the links between those users' pages for a single-time snapshot from September 2005.\footnote{We have posted the data at \url{http://people.maths.ox.ac.uk/~porterm/data/facebook5.zip}.} Similar snapshots of Facebook data from 10 Texas universities were analyzed recently in Ref.~\cite{oldboy}, and a snapshot from ``a diverse private college in the Northeast U.S.'' was studied in Ref.~\cite{lewis}.  Other studies of Facebook have typically obtained data either through surveys \cite{boydell} or through various forms of automated sampling \cite{ucifacebook}, thereby containing missing nodes and links that can strongly impact the resulting graph structures and analyses.

We consider only ties between people at the same institution, which yields five separate realizations of university social networks and allows us to compare the structures at different institutions.  Our study includes a small technical institute (California Institute of Technology [Caltech]), a pair of private universities (Georgetown University and Princeton University), and a pair of large state universities (University of Oklahoma and University of North Carolina at Chapel Hill [UNC]).

\begin{figure}
\centerline{\includegraphics[height=.1565\textwidth,width=.22\textwidth]{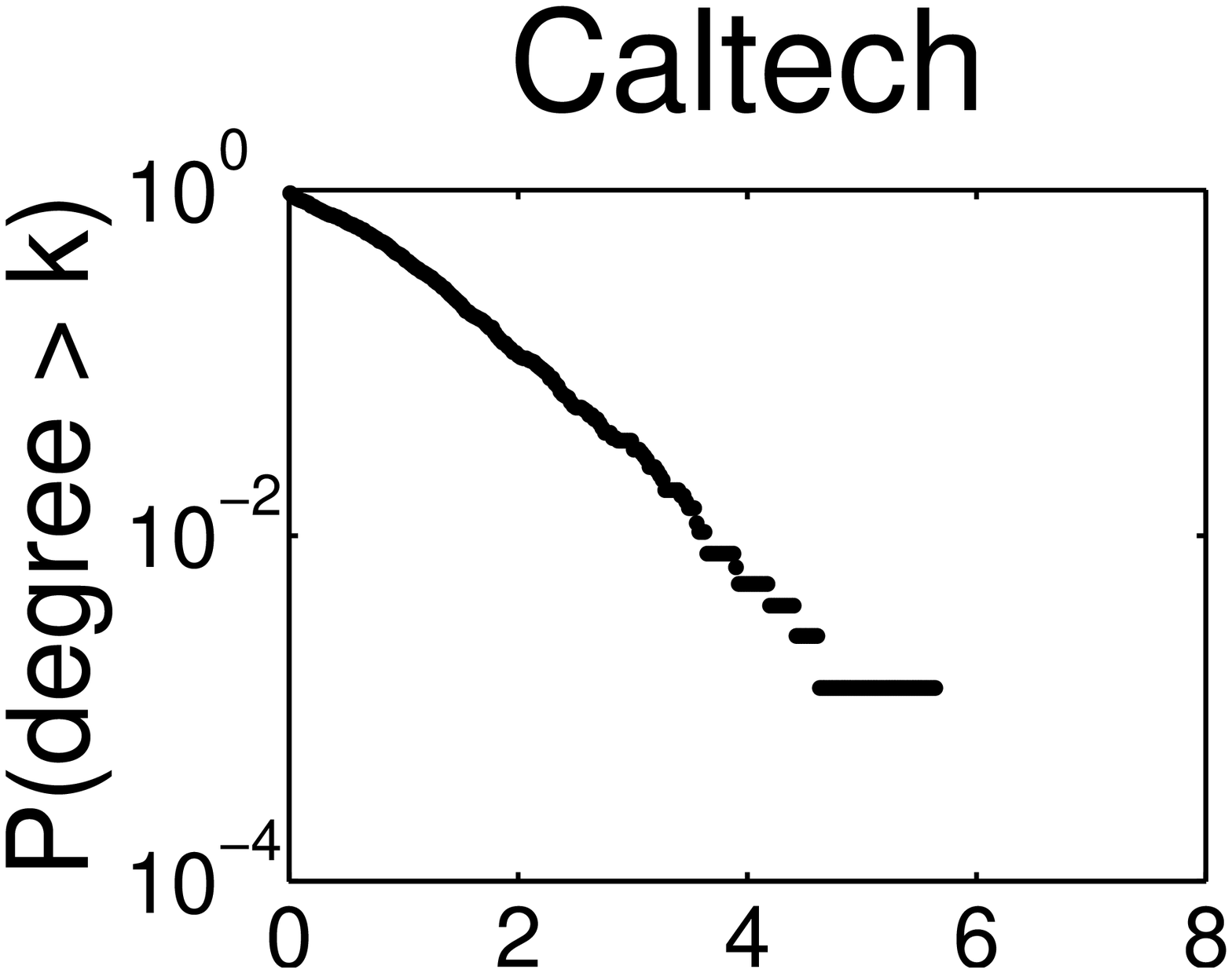}
\includegraphics[height=.1565\textwidth]{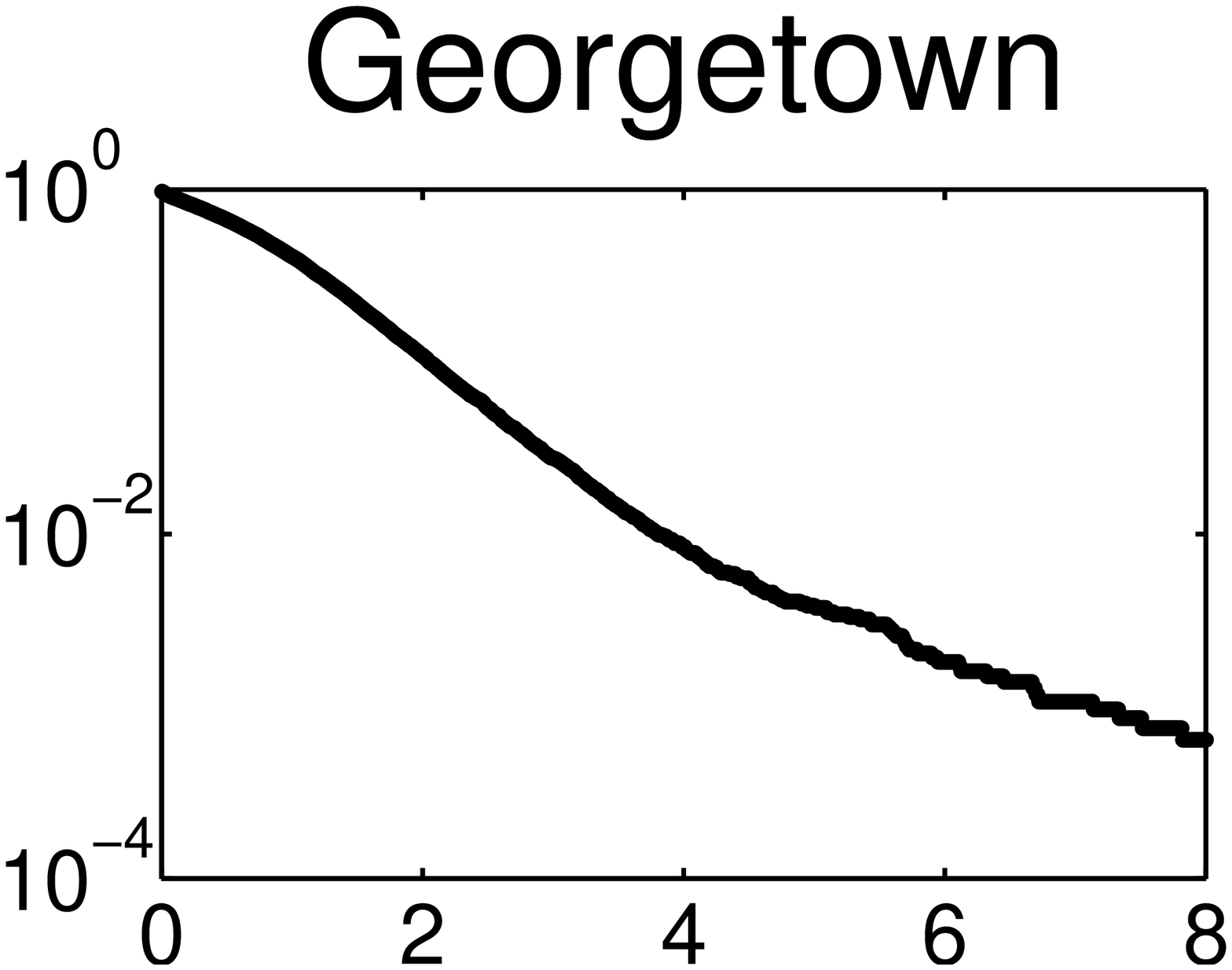}
\includegraphics[height=.1565\textwidth]{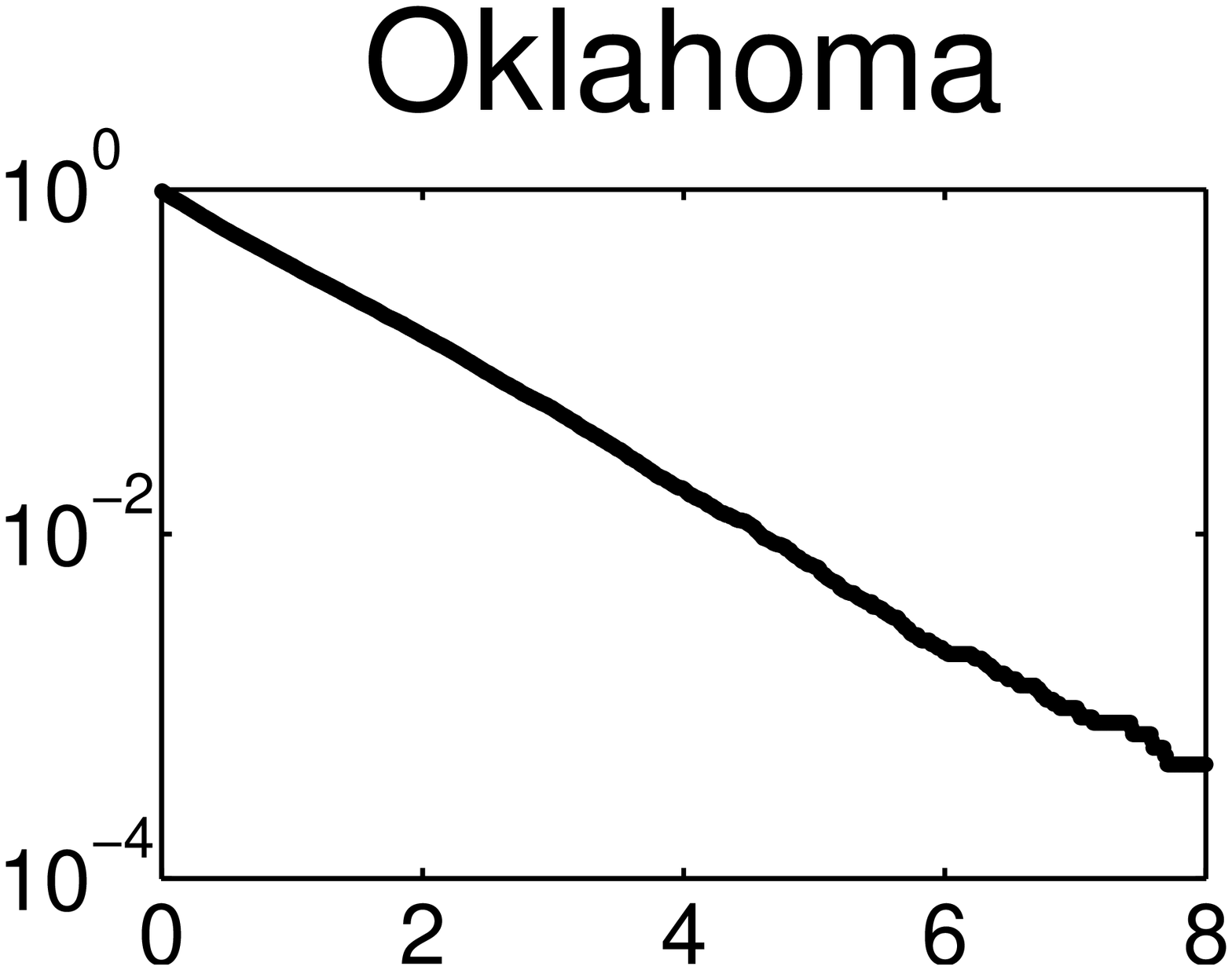}
\includegraphics[height=.1565\textwidth]{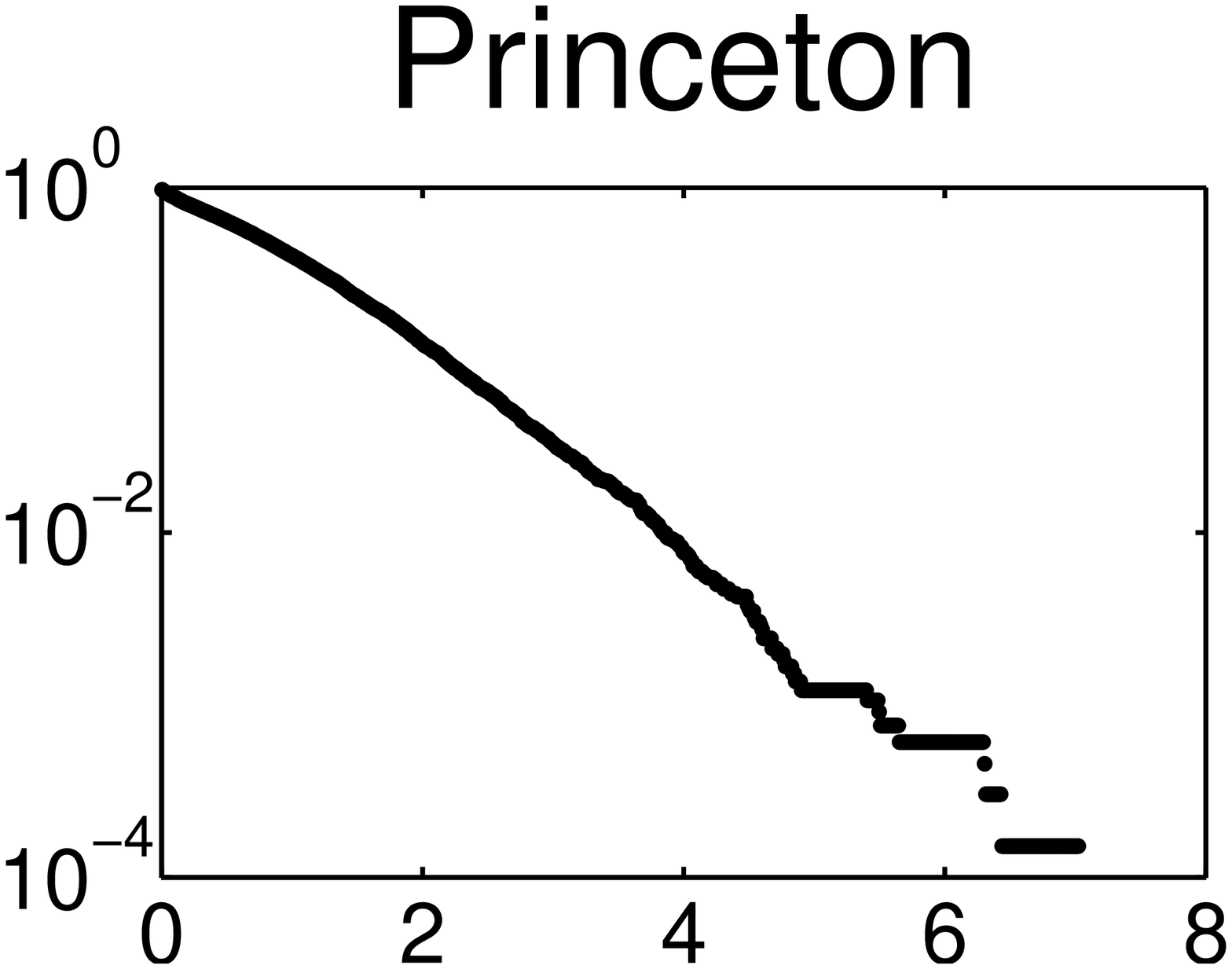}
\includegraphics[height=.1565\textwidth]{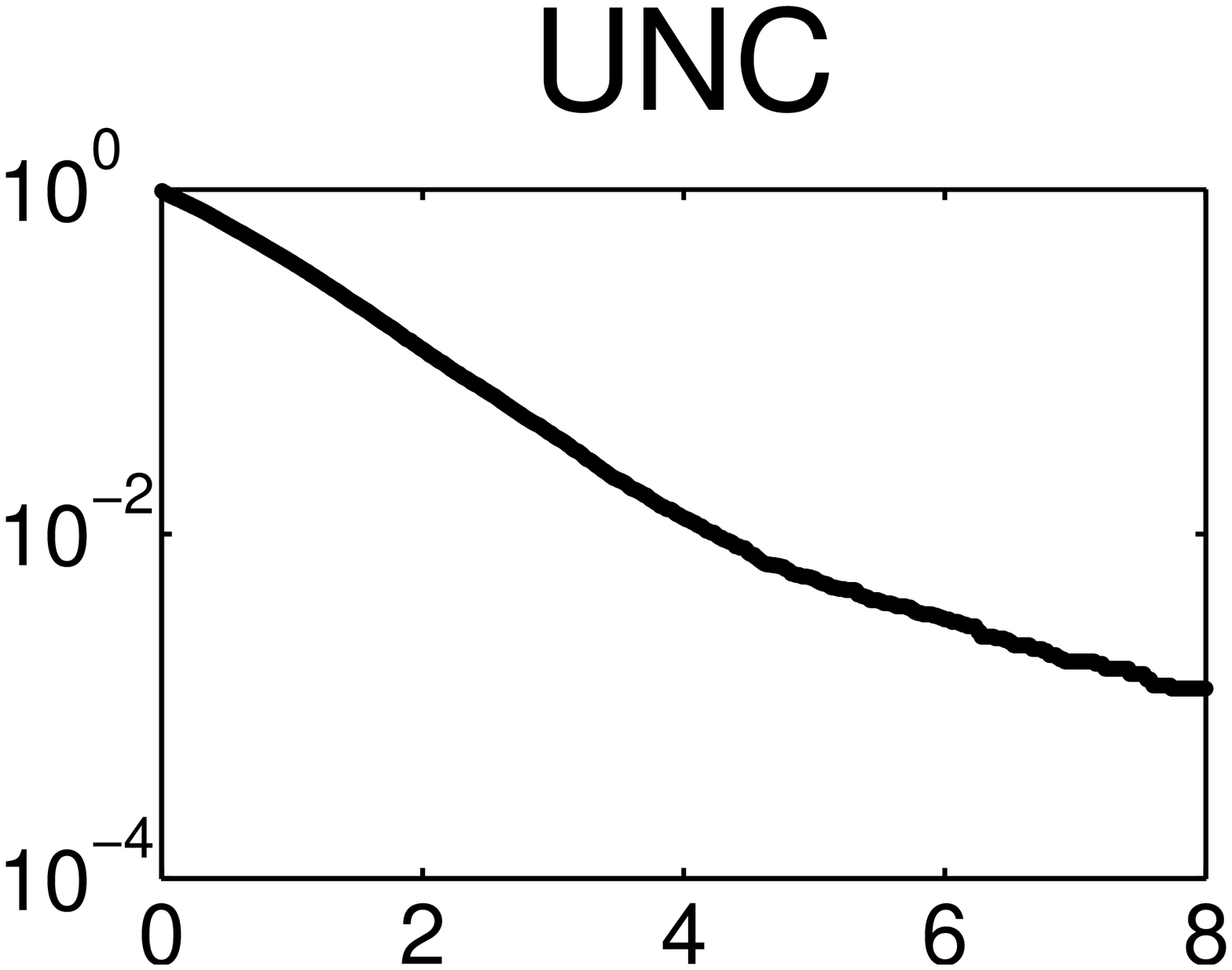}}
\centerline{\includegraphics[height=.16\textwidth,width=.22\textwidth]{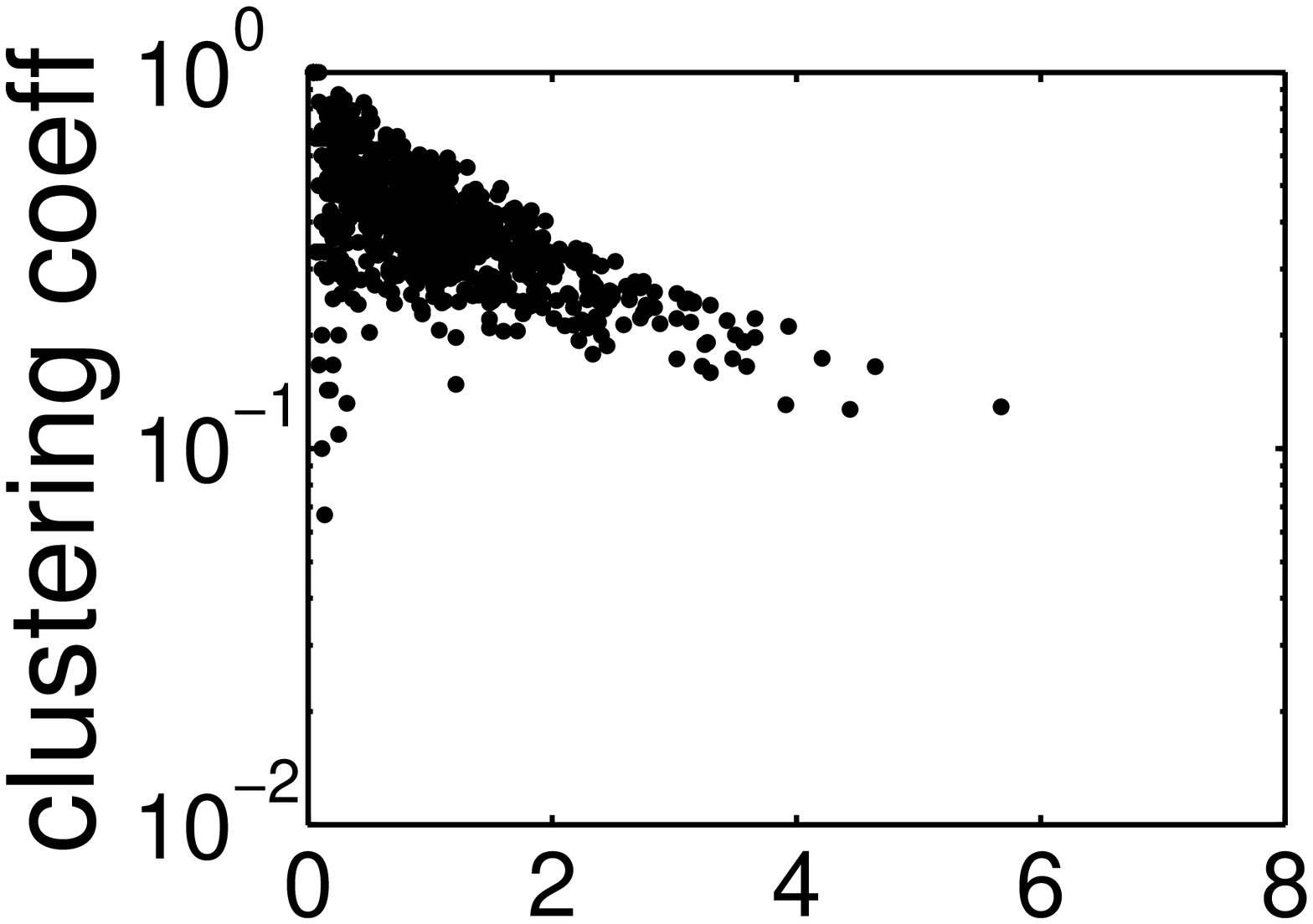}
\includegraphics[height=.16\textwidth]{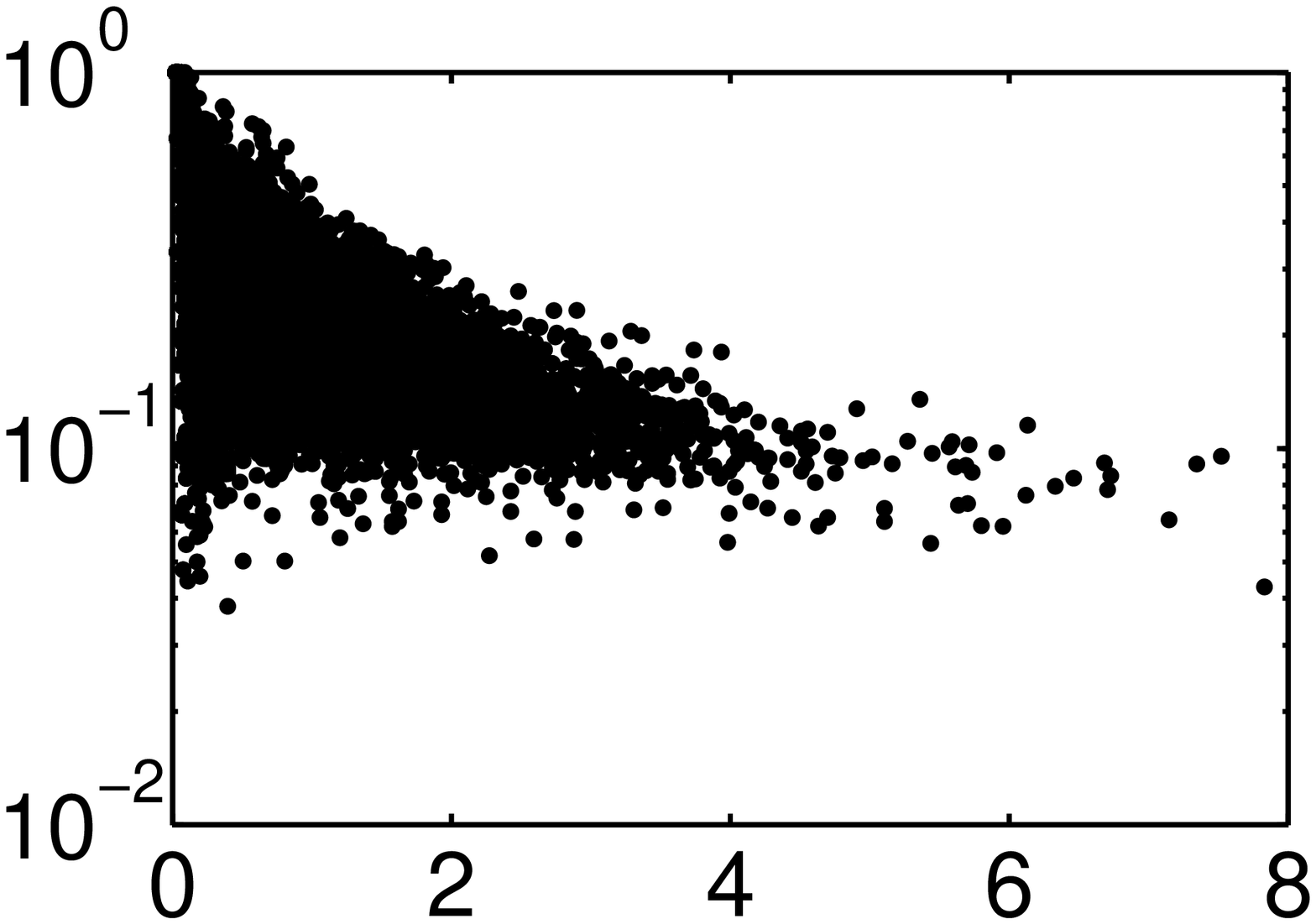}
\includegraphics[height=.16\textwidth]{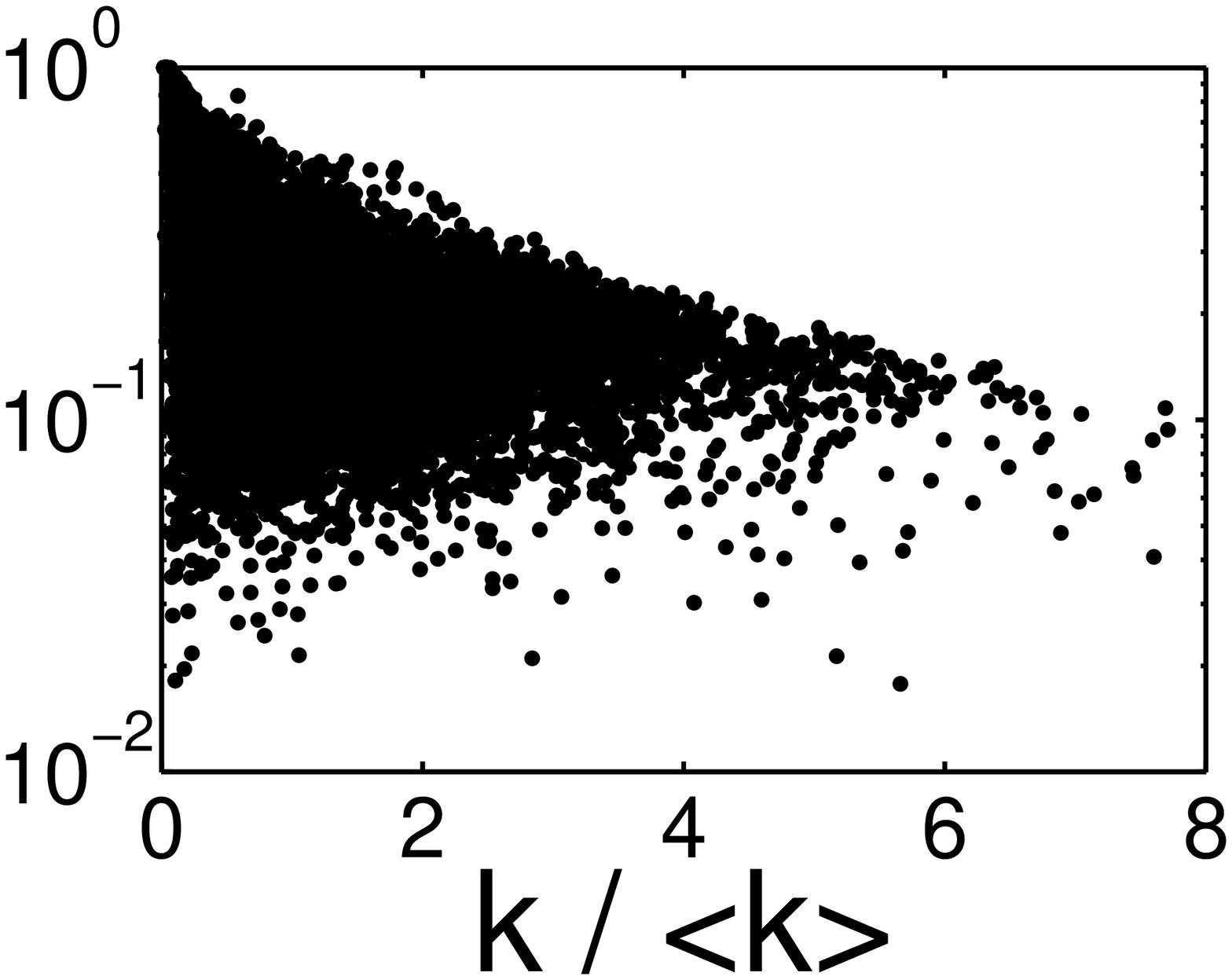}
\includegraphics[height=.16\textwidth]{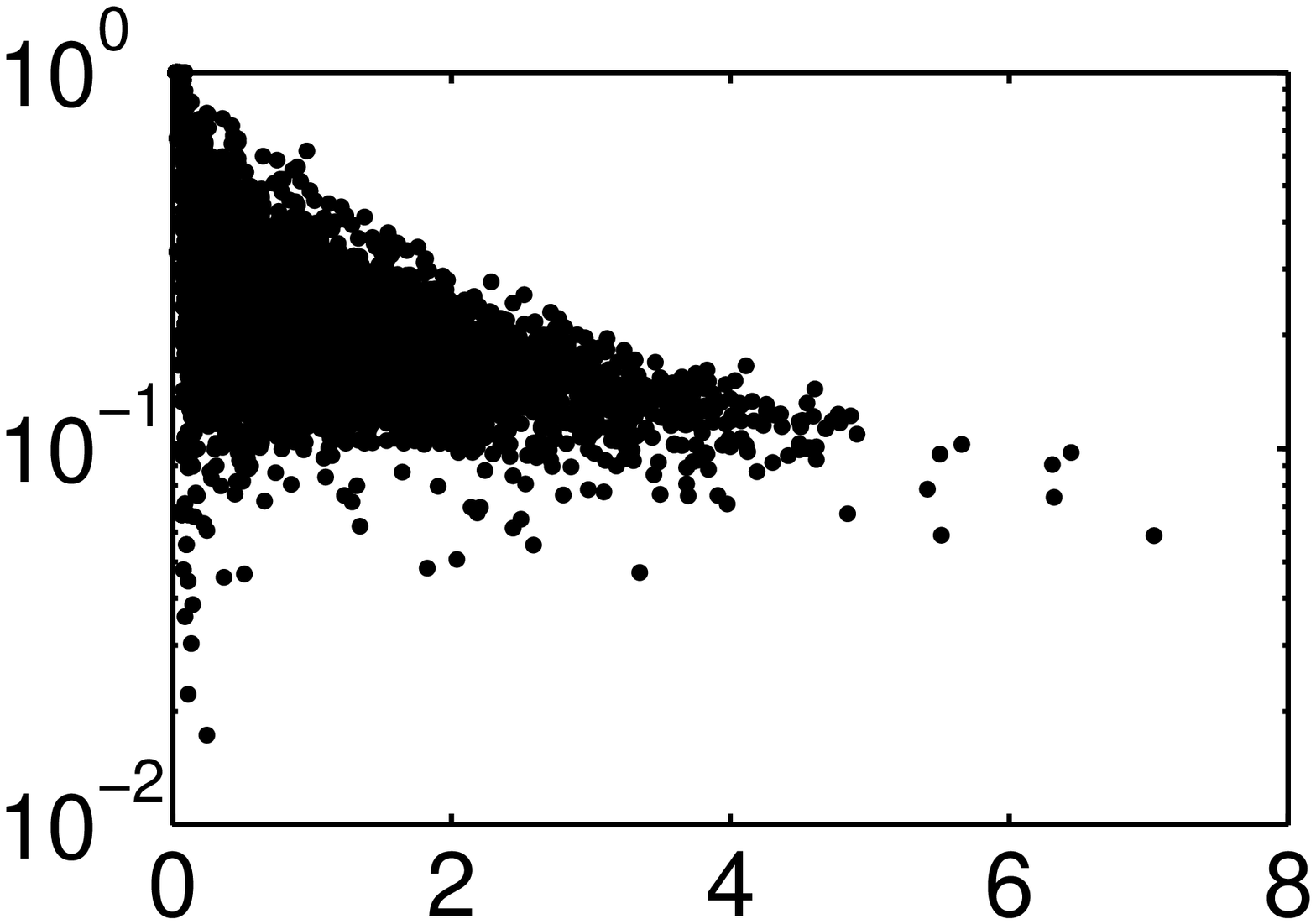}
\includegraphics[height=.16\textwidth]{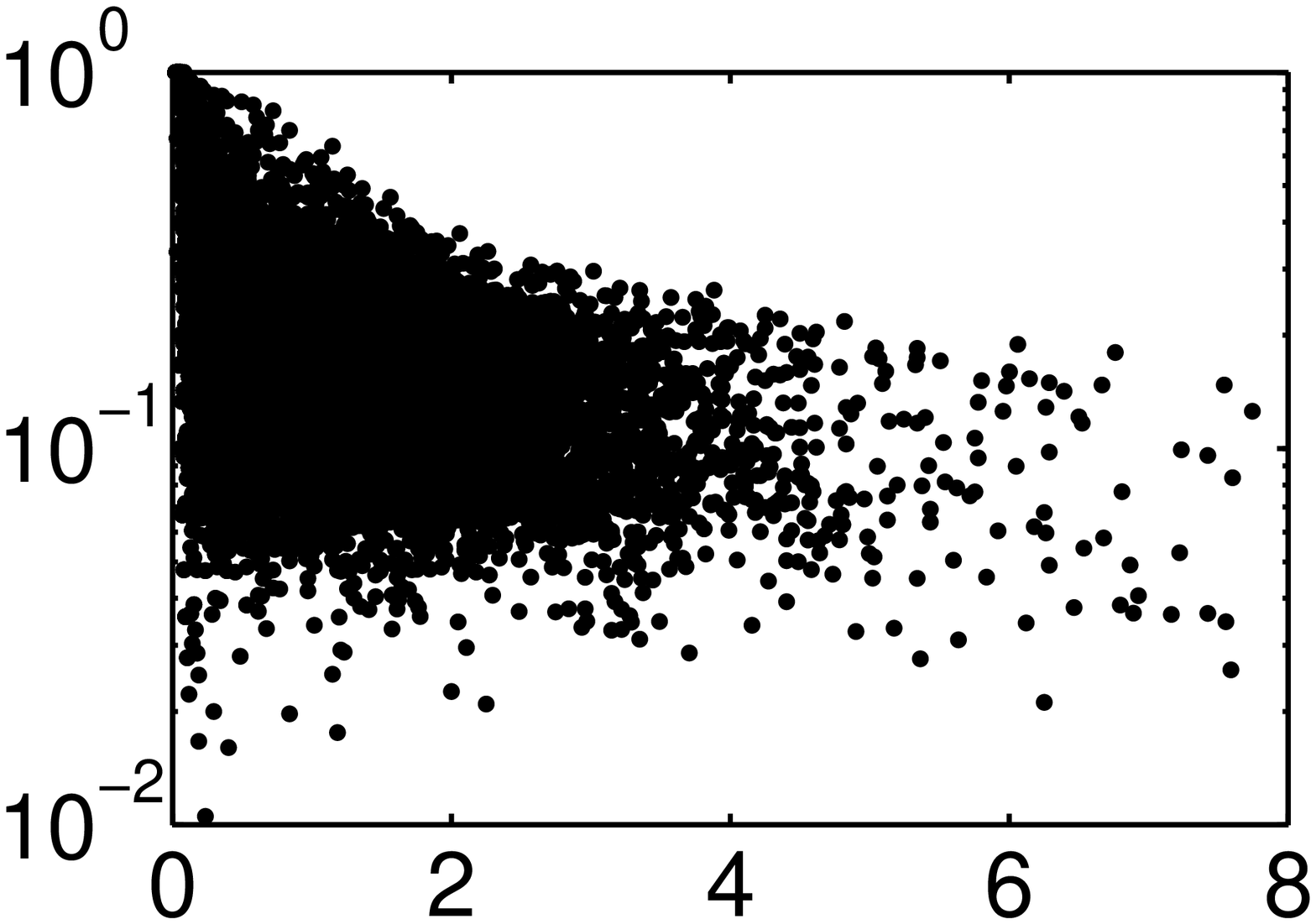}}
\caption{Cumulative degree distributions (top panels) and local clustering coefficients (bottom panels) for the five university networks.  We employ semilogarithmic coordinates.  The horizontal axes give the degree relative to the mean degree $\langle k \rangle$, and we only display data for $k/\langle k \rangle \leq 8$ to provide common axes for all universities. 
}
\label{sumsum}
\end{figure}

We summarize basic properties of the university networks in Fig.~\ref{sumsum} and Table \ref{table:size}.  See \cite{Newman2003,newman2010} and references therein for discussions of the measures that we use in this section.  Although our focus in this paper is community structure, we remark that even these simple network characteristics can yield insights about Facebook networks.  The mean degrees tend to increase with network size, potentially indicating that broader institutional use begets greater personal use (though this trend is clearly strongly influenced by the Caltech data). The degree distributions of these institutions (plotted in the top panels of Fig.~\ref{sumsum}) have heavy tails compared to random graphs. In particular, the degree distributions appear to be approximately exponential.
Although the mechanisms driving such distributions are impossible to ascertain without longitudinal data, the roughly exponential form of the degree distribution both above and below the mean degree potentially indicates a wide range in the willingness to participate (i.e., to add online friends) among Facebook users. 

The bottom panels of Fig.~\ref{sumsum} compare node degree versus clustering coefficient,
\begin{align*}
	C_i = \frac{\mbox{number of pairs of neighbors of node } i \mbox{ that are connected} }{\mbox{number of pairs of neighbors of node } i}\,.
\end{align*}
We note that even heavy users have much larger local clustering than that expected at random (e.g., when compared with the total graph densities).  In Table \ref{table:size}, we provide the mean clustering coefficient and the transitivity for each network, given by the fraction of connected triples in the network that are fully connected triangles. Both measures of local clustering are much larger at Caltech than they are at the other institutions. It is of course not surprising that we observe large transitivities in social networks such as the Facebook networks.  Nevertheless, as we have shown recently in Ref.~\cite{melnik}, tree-based theories of various dynamical processes appear to be valid for Facebook networks (despite their high clustering, implying that they are most definitely not locally tree-like) because they are ``sufficiently small" worlds, in that the mean distance between nodes is close to the expected value obtained in random networks with the same joint degree-degree distributions.

\begin{table}
\centerline{\footnotesize
\begin{tabular}{|r|c|c|c|c|c|}\hline
Institution & Caltech & Georgetown & Oklahoma  & Princeton & UNC\\ \hline
Nodes & 1099 &12195 & 24110 & 8555 &  24780\\
Connected Nodes & 762  & 9388 & 17420 & 6575 &  18158\\
Connected Edges & 16651 & 425619 & 892524 & 293307 &  766796 \\
Mean Degree & 43.7 & 90.7 & 102.5 & 89.2 & 84.5\\
Mean Clustering Coeff. & 0.4091 & 0.2249 & 0.2297 & 0.2372 & 0.2020\\
Transitivity & 0.2913 & 0.1485 & 0.1587 & 0.1639 & 0.1156 \\ \hline
Degree Assortativity & $-0.0662$ & 0.0753 & 0.0737 & 0.0910 & 6.6$\times 10^{-5}$ \\
Gender Assortativity & 0.0540 & 0.0145 & 0.1118 & 0.0650 & 0.0598\\
Major Assortativity & 0.0382 & 0.0439 & 0.0412 & 0.0474 & 0.0511\\
Dormitory Assortativity & 0.4486 & 0.1725 & 0.4033 & 0.0872 & 0.2024\\
Year Assortativity & 0.2694 & 0.5575 & 0.2923 & 0.4947 & 0.3964\\
High School Assortativity & 0.0021 & 0.0237 & 0.1583 & 0.0197 & 0.1342\\ \hline
Number of Communities & 12 & 33 & 5 & 12 & 5 \\
Modularity & 0.4003 & 0.4801 & 0.3869 & 0.4527 & 0.4274 \\ \hline
\end{tabular}}
\caption{Basic characteristics of the largest connected components of the five Facebook networks that we study: the total number of nodes in the original data, numbers of nodes and edges in the largest connected component, mean degree, mean clustering coefficient, transitivity (fraction of transitive triples), assortativities (by degree, gender, major, dormitory, class year, and high school), number of communities detected, and the modularity of the resulting graph partition. In calculating the assortativities, we ignored nodes for which the corresponding demographic characteristic is missing (i.e., the ``pairwise removal'' protocal that we discuss in Section \ref{sec:examples}).  We treat class year as a categorical variable here, and we calculate degree assortativity as a correlation coefficient \cite{Newman2003,newman2010}.
}
\label{table:size}
\end{table}

The data also includes limited demographic information provided by users on their individual pages: gender, class year, and data fields that represent (using anonymous numerical identifiers) high school, major, and dormitory residence (or ``House'' at Caltech).  In situations in which individuals elected not to volunteer a demographic characteristic, we use an additional ``Missing'' label. These characteristics allow us to make comparisons between different universities, under the assumption (per the discussion in Ref.~\cite{boydell}) that the communities and other elements of structural organization in Facebook networks reflect (even if imperfectly) the social communities and organization of the offline networks on which they're based.

For instance, at the level of individual ties, the tendency for users to be friends with other users who have similar characteristics can be quantified by the assortativity of the links relative to that characteristic. Degree assortativity (or degree correlation) can be calculated as the Pearson correlation coefficient of the degrees at either ends of the edges. Although many social networks tend to be positively assortative with respect to degree, we find that the degree assortativity is negative for Caltech and is very small for UNC.  A general measure of scalar assortativity relative to a categorical variable is given by 
\begin{equation}
	r = \frac{\mbox{tr} ({\bf e}) - \| {\bf e}^2 \|}{1 - \| {\bf e} ^2 \|} \in [-1,1] \,,
\end{equation}
where ${\bf e} = {\bf E }/\| {\bf E} \|$ is the normalized mixing matrix, the elements $E_{ij}$ give the number of edges in the network that connect a node of type $i$ (e.g., a person with a given major) to a node of type $j$, and the entry-wise matrix $1$-norm $\| {\bf E } \|$ is equal to the sum of all entries of ${\bf E}$.  Comparing assortativities for various categories shows, for example, that assortativity by dormitory and class year (treated as a categorical variable) are high for all five institutions; assortativities by major are low for all five institutions; and assortativities by high school and gender are less consistent across institutions.  The relative sizes of the different assortativities also vary across institutions, which is similar to what we will see below with communities.  Going beyond this measure of local assortativity by characteristics, our major focus for this article is on the  organization of the communities of these five Facebook networks based on these various categories.  We discuss this in detail in Section \ref{sec:examples}.


\section{Facebook Communities} \label{sec:examples}

We algorithmically identify a set of communities in the largest connected component of each institution's network using a modified version of Newman's leading-eigenvector method \cite{newman2006pre} in conjunction with subsequent Kernighan-Lin node-swapping steps \cite{kl}. We compare the communities to partitions obtained by grouping users according to each of the self-identified characteristics: major, class year, high school, and dormitory/House.

We first revisit Caltech's community structure, which we previously examined visually in Fig.~\ref{fig:Caltech}.  The partition of the largest connected component into 12 communities (which has modularity $Q \doteq 0.4003$) exhibits a strong correlation with House affiliation.  To investigate this quantitatively, we calculate the similarity coefficients of this partition versus each partition constructed using one of the four available user characteristics (see Table \ref{table:caltech_S}).  The raw $S_i$ values appear to be insufficient to the task of comparing these communities.  Specifically, the ordering of the correlation strengths with the different demographics is not consistent across pair-counting indices, even among those we know are linear transformations of one another.  Additionally, although there is agreement that the correlation with House is strongest, the $S_i$ values differ wildly in how much they set apart the House correlation, with $S_\mathrm{R}$ and $S_\mathrm{M}$ seemingly indicating that the correlation with House is only marginally stronger than that with high school even though Caltech contains very few students at one time that come from the same high school.

These apparent disagreements in interpretation across $S_i$ values occur even though we know that their corresponding $p$-values in the (unobtained) random distributions are identical.  While we cannot directly calculate those $p$-values, the $z$-scores for each (see Section \ref{subsec:zscores}) in Table \ref{table:caltech_S} indicate that the correlation with high school is the only one of the four demographic characteristics that is not statistically significant. We note that the ordering of the VI scores in Table \ref{table:caltech_S} is consistent with that of the $z$-scores but recall that such agreement of ordering is not consistently observed in Fig.~\ref{fig:ZvZR}.
The $z$-scores provide a consistent interpretation of the roles of the four characteristics in this Caltech data: House is most important, followed distantly by year and major (in descending order), with no significant correlation with high school.  Because of the close agreement between the $z_\mathrm{J}$, $z_\mathrm{M}$, and $z_\mathrm{R}$ scores in Fig.~\ref{fig:ZvZR} and Table \ref{table:caltech_S}, we henceforth restrict attention to the analytically-obtained $z_\mathrm{R}$ values.

\begin{table}
\centerline{\footnotesize
\begin{tabular}{r|c|c|c|c|c|c|c|c|c|c|}
& $S_\mathrm{AR}$ & $S_\mathrm{FM}$ & $S_\Gamma$ & $S_\mathrm{J}$ & $S_\mathrm{M}$ & $S_\mathrm{R}$ & VI
& $z_\mathrm{J}$ & $z_\mathrm{M}$ & $z_\mathrm{R}$ \\ \hline
``Major'' & 0.0063 & 0.1195 & 0.0070 & 0.0576 & 1.1238 & 0.7785 & 4.3149 & 3.96 & 3.95 & 3.96\\
``House'' & 0.3762 & 0.4742 & 0.3829 & 0.3056 & 0.9578 & 0.8391 & 1.9275 & 249 & 226 & 198\\
``Year'' & 0.0080 & 0.1766 & 0.0080 & 0.0968 & 1.2637 & 0.7199 & 3.5191 & 6.84 & 6.82 & 6.73\\
``H.S.'' & 0.0085 & 0.0833 & 0.0129 & 0.0301 & 1.0484 & 0.8072 & 4.7268 & -0.55 & -0.55 & -0.55\\
\hline
\end{tabular}}
\caption{Similarity coefficients (Adjusted Rand, Fowlkes-Mallows, $\Gamma$, Jaccard, Minkowski, and Rand), variation of information, and similarity $z$-scores for comparing a 12-community partition of the Caltech data versus a partition constructed using each of the four self-identified user characteristics.}
\label{table:caltech_S}
\end{table}

Before concluding our discussion of Caltech, we acknowledge the potentially important effects of missing demographic data, as a significant number of users did not volunteer an affiliation (as indicated in Table \ref{table:sizemissing} and by the purple wedges of Fig.~\ref{fig:Caltech}). One can approach the issue of missing data using sophisticated tools such as multiple imputation, likelihood, or weighting methods \cite{muchado}.  A simpler approach is to investigate the effects on the measured correlations by various restrictions of the data. We consider three such protocols: inclusion, pairwise removal, and listwise removal. Inclusion, which we use in Table \ref{table:caltech_S}, treats the missing labels like any other category, erroneously grouping all such users together in the demographic partition.  We apply pairwise removal separately for each demographic comparison with the community structure.  In terms of a contingency table of $r$ demographic rows and $c$ community columns, this amounts to a deletion of the row corresponding to ``Missing.''  Listwise removal restricts the comparisons to the subset of users who volunteered all four of the studied demographic characteristics. We stress that these protocols do not affect the community assignments, which we obtained using the complete network data. Other restrictions or combinations of this data (such as single-gender restrictions) can also be fruitfully explored,
but such investigations are beyond the scope of the present article.


\begin{table}
\centerline{\footnotesize
\begin{tabular}{r|c|c|c|c|c|c|}
& Connected & Indicated & Indicated & Indicated & Indicated & Indicated\\
& Users & Major & Dorm/House & Year & High School & All\\ \hline
Caltech & 762 & 687 & 594 & 651 & 633 & 499\\
Georgetown & 9388 & 7510 & 6594 & 8374 & 7562 & 4774\\
Oklahoma & 17420 & 15779 & 7203 & 13732 & 14998 & 5510\\
Princeton & 6575 & 4940 & 4355 & 5801 & 5214 & 2920\\
UNC & 18158 & 15492 & 8989 & 15883 & 15414 & 6719\\ \hline
\end{tabular}}
\caption{Numbers of nodes of each data set used in the different protocols for treating missing data.} \label{table:sizemissing}
\end{table}

In Table \ref{table:zscores}, we present the $z_\mathrm{R}$-scores for all four community-demographic comparisons using each of the three missing data protocols at the five universities we study. We caution that because of network-size effects (reflecting the different numbers of nodes in different examples), $z$-score values cannot typically be directly compared across institutions. Accordingly, our primary conclusions are about the statistical significances and rank orderings of the demographic correlations separately in each university.  Our previous conclusions about the Caltech community structure remain largely consistent across all three missing data protocols: House is most strongly correlated with the communities, followed distantly by year and major (in descending order), with no statistically significant correlation with high school. While House remains strongly correlated with communities in all three protocols, the correlation with year and major appears to be only marginally statistically significant in the analysis with listwise removal.

\begin{table}
\centerline{\small
\begin{tabular}{r|c|c|c|c|c|}
& Caltech & Georgetown & Oklahoma & Princeton & UNC \\ \hline
{\bf Inclusion:} \hfill ``Major'' & 3.962 & 5.885 & 3.799 & 15.03 & 8.044 \\
``Dorm/House'' & 200.8 & 148.8 & 71.00 & 58.26 & 113.0 \\
``Year'' & 6.727 & 1543 & 206.7 & 1058 & 778.2 \\
``High School'' & -0.553 & 26.13 & 18.50 & 15.62 & 15.93 \\
\hline
{\bf Pairwise:} \hfill ``Major'' & 4.051 & 16.00 & 16.44 & 9.968 & 5.700 \\
``Dorm/House'' & 285.3 & 212.9 & 186.9 & 147.2 & 93.34 \\
``Year'' & 5.389 & 1837 & 286.1 & 1270 & 889.1 \\
``High School'' & 0.7695 & 4.247 & 22.54 & 2.888 & 37.22 \\
\hline
{\bf Listwise:} \hfill ``Major'' & 2.235 & 15.23 & 26.10 & 10.07 & 13.90 \\
``Dorm/House'' & 248.9 & 221.5 & 159.9 & 116.5 & 90.50 \\
``Year'' & 2.644 & 1913 & 251.2 & 997.3 & 475.7 \\
``High School'' & 0.3063 & 1.228 & 13.69 & 2.415 & 21.12 \\
\hline
\end{tabular}}
\caption{Analytically-obtained $z_\mathrm{R}$-scores for comparing the algorithmically-identified communities of Facebook networks versus user characteristics. Cases where users did not volunteer demographic characteristics are treated by three protocols: inclusion, pairwise removal, and listwise removal.
}
\label{table:zscores}
\end{table}

\begin{figure}[ht]
    \centerline{
        \includegraphics[width=0.4\textwidth]{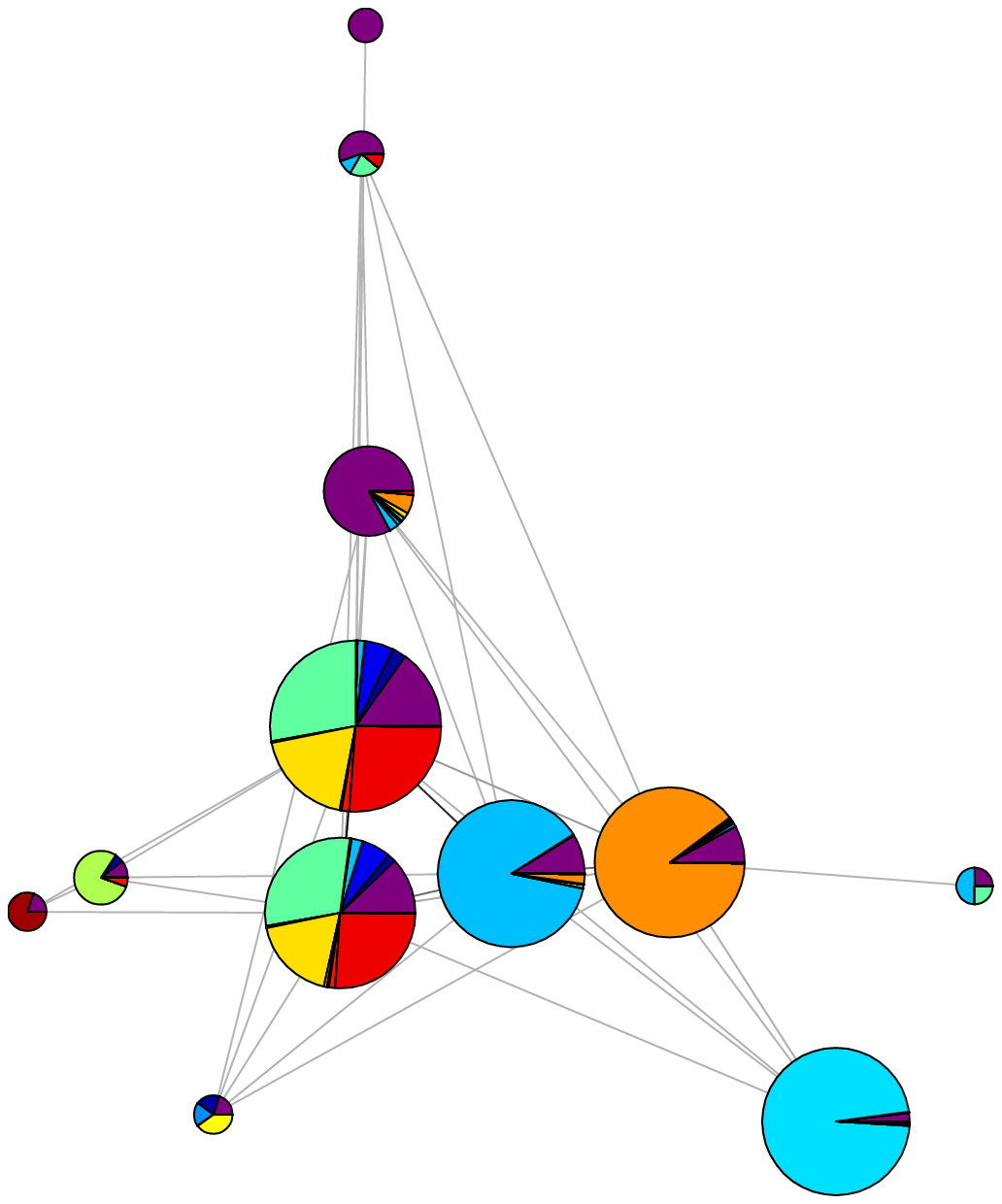}\qquad
        \includegraphics[width=0.4\textwidth]{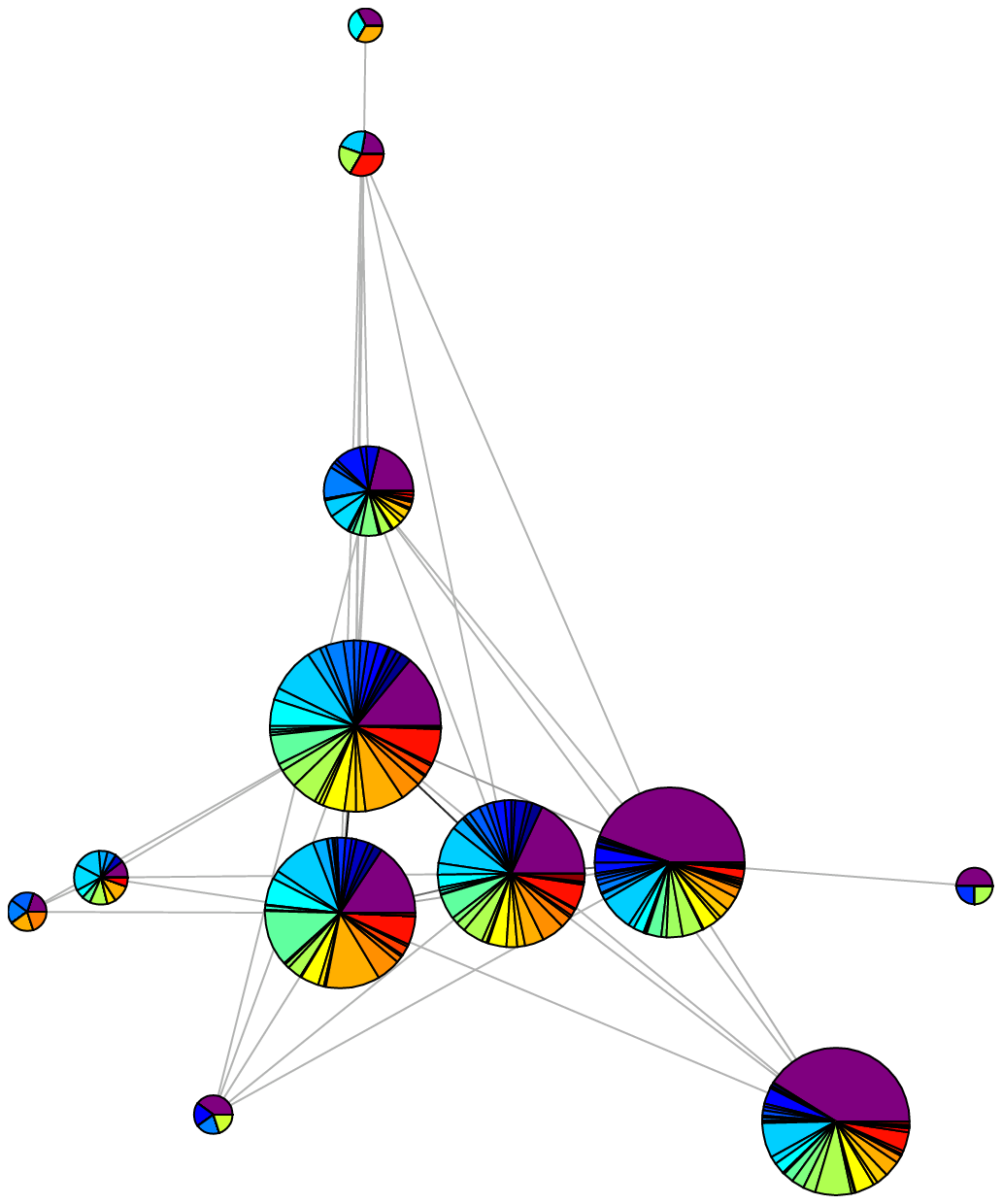}
        }
    \caption{[Color] Pie-charts of Princeton, colored by (Left) class year and (Right) major. (As before, purple slices correspond to people who did not identify the relevant characteristic.)}
    \label{fig:PrincetonbyYearMajor}
\end{figure}

In contrast with Caltech, the communities at each of the other four institutions that we study correlate primarily with class year (see Table \ref{table:zscores}). Moreover, these correlations are not as dominant as House is at Caltech, as each of the four characteristics possess statistically significant correlations with the community structures at the other four institutions (except high school in listwise removal at Georgetown).  We show the 12 communities identified at Princeton colored both by class year and by major in Fig.~\ref{fig:PrincetonbyYearMajor}. Compared with the strong correlation between communities and House affiliation at Caltech, these visual depictions of the Princeton communities do not seem to indicate as strong a correlation with year despite the very large corresponding $z_\mathrm{R}$ (which again cautions against direct comparison of $z_\mathrm{R}$ values in networks of different sizes).  We remark that the size of the Princeton data set, with over 8500 nodes (6575 in the largest connected component) is disproportionately large relative to the institution's size; this is presumably a result of the relatively early Facebook adoption there.

The $z$-scores in Table \ref{table:zscores} reveal that Princeton students break up into communities primarily according to class year (among the four demographic categories available to us), and dormitory gives the second highest correlation. While major is also significant, the correlation with high school appears to be only marginally significant in protocols that remove missing data.  One can draw similar conclusions about Georgetown from Table \ref{table:zscores}; the only qualitative difference is the possible lack of significance of high school at Georgetown (as compared to the marginal significance at Princeton) that is suggested by the more stringent missing-data protocols.

Similarly, the $z$-scores calculated for the UNC network partitioned into 5 communities suggest that class year is the primary organizing characteristic and that dormitory residence is also prominent. High school and major have smaller but significant positive correlations with the community structure. The other large state university that we consider is the University of Oklahoma, which is also partitioned into 5 communities. Like UNC, the dominant correlation of the Oklahoma communities is with year, the secondary correlation is with dormitory, and both high school and major have statistically significant correlations. Unlike UNC, however, the disparity between the correlations with year and with dormitory do not appear to be as wide at Oklahoma. In contrast to Princeton and Georgetown, communities at both UNC and Oklahoma maintain unquestionably significant correlations with high school in both missing-data protocols.

We close this section by cautioning about interpretations of conclusions drawn from the numbers in Table~\ref{table:zscores}, even though they indicate some interesting differences among the institutions that we studied. In particular, one should of course be careful about how such numbers might be influenced by our methodologies. Although we have provided three different protocols for handling missing data, other effects might be similarly worthy of study. For instance, one should be wary of the possible influence of the selected definition of ``community'' and the method of its detection. There are numerous definitions and methods available (again see Refs.~\cite{ourreview,santobig}), and a more definitive analysis of the connections between communities and characteristics in such networks should more fully explore multiple notions of community, possibly hierarchical structures, and communities at different resolutions.

As a simple example of comparing results from different community-detection methods, we compare the 12-community Caltech partition with that obtained for a 7-community partition (with $Q \doteq 0.3594$), which we obtained using a simpler spectral modularity-optimization implementation. Despite the necessarily different details of these two community structures, the qualitative conclusions from the two partitions are the same: House provides the dominant correlation, followed distantly by year and major, and there is again no significant correlation high school.  Applying this same ``weaker" (in the sense of consistently resulting in partitions of lower modularity) community-detection implementation to the other four institutions also typically agrees with the results that we report above: Year has the strongest correlation with communities and is followed by dormitory.  The role of high school appears to be more pronounced in these lower-modularity partitions, as one obtains statistically significant correlations with the communities at Georgetown and Princeton and even stronger correlations with the communities at UNC and Oklahoma.

We also stress the difference between causation and correlation.  In this paper, we have examined \emph{correlations}.  As discussed in the sociological literature on SNSs (see \cite{boydell} and references therein), it is obviously very interesting and important to attempt to discern which common characteristics have resulted from friendships and which ones might perhaps influence the formation of friendships.  In terms of the individual characteristics discussed above, high school and class year are known prior to the formation of these Facebook links, so one would expect those particular correlations to also indicate how some friendships might have formed.  Common residences and majors, on the other hand, can both encourage new friendships and arise because of them.  We note, finally, that SNS friendships provide only a surrogate for offline ones, so that one can also expect to find some differences between the community structures of Facebook networks and the real-life networks that they imperfectly represent \cite{boydell}.


\section{Conclusions} \label{sec:discussion}

We have demonstrated that analysis of community structure is useful for studying the online social networks of universities and inferring interesting insights about the prominent driving forces of community development in their corresponding offline social networks.  We investigated various measures for comparing algorithmically-identified communities in Facebook networks with those obtained by grouping individuals according to self-identified characteristics. We found that $z$-scores of pair-counting indices provide an immediate (though not quantitatively perfect) interpretation about the likelihood that such values might arise at random, indicating significant correlations between the algorithmically-identified communities and multiple self-identified characteristics. Such calculations indicate that the organizational structure at Caltech, which depends very strongly on House affiliation, is starkly different from those of the other universities that we studied. The observed heterogeneity in the communities, even at an institution like Caltech whose social structure seems to be mostly dominated by a single feature (House affiliation), underscores the important point that networks typically have multiple organizational forces \cite{ourreview}. We hope that our work leads to a wider comparative study that might increase understanding about the different factors that drive the social organization of universities.  The present paper attempts to provide foundational steps for such comparative investigations by conveying a meaningful methodology.


\section*{Acknowledgements}

We thank Adam D'Angelo and Facebook for providing the data used in this study.  We also acknowledge Skye Bender-de Moll, Danah Boyd, Barry Cipra, Barbara Entwisle, Katie Faust, Avi Feller, Dan Fenn, James Gleeson, Sandra Gonz\'{a}lez-Bail\'{o}n, Justin Howell, Nick Jones, Franziska Klingner, Marco van der Leij, Tom Maccarone, Jim Moody, Mark Newman, Andy Shaindlin, and Ashton Verdery for useful discussions.  We are especially indebted to Aaron Clauset and James Fowler for thorough readings of a draft of this manuscript and to Christina Frost for developing some of the graph visualizations we used.\footnote{The code is available at \url{http://netwiki.amath.unc.edu/VisComms}.}  ALT was funded by the NSF through the UNC AGEP (NSF HRD-0450099) and by the UNC ECHO program. EDK's primary contributions to this project were funded by Caltech's Summer Undergraduate Research Fellowship (SURF) program. PJM was funded by the NSF (DMS-0645369) and by start-up funds provided by the Institute for Advanced Materials, Nanoscience \& Technology and the Department of Mathematics at the University of North Carolina at Chapel Hill.  MAP did some work on this project while a member of the Center for the Physics of Information at Caltech and also acknowledges a research award (\#220020177) from the James S. McDonnell Foundation.

\bibliographystyle{siam}

\end{document}